\documentclass[prl,amsmath,amssymb]{revtex4}
 
\usepackage{graphicx}
\usepackage{subfigure}
\usepackage{dcolumn}
\usepackage{array}
\usepackage{bm}
\usepackage{natbib} 

\begin{document}
 
\title{\bf{Integration of first-principles methods and crystallographic
database searches for new ferroelectrics: Strategies and explorations}\\[11pt] }
\author{Joseph W. Bennett and Karin M. Rabe}
\address{ Department of Physics and
Astronomy\\ Rutgers University, Piscataway, NJ 08854} \date{\today}

\begin{abstract}
In this concept paper, the development of strategies for the
integration of first-principles methods with crystallographic database
mining for the discovery and design of novel ferroelectric materials
is discussed, drawing on the results and experience derived from
exploratory investigations on three different systems: (1) the double
perovskite Sr(Sb$_{1/2}$Mn$_{1/2}$)O$_3$ as a candidate semiconducting
ferroelectric; (2) polar derivatives of schafarzikite $M$Sb$_2$O$_4$;
and (3) ferroelectric semiconductors with formula
$M_2$P$_2$(S,Se)$_6$. A variety of avenues for further research and
investigation are suggested, including automated structure type
classification, low-symmetry improper ferroelectrics, and
high-throughput first-principles searches for additional
representatives of structural families with desirable functional
properties.
\end{abstract}


\maketitle

\section{\label{sec:level1} Introduction}

One of the central challenges in modern materials science is the
discovery and design of functional materials that can be readily
incorporated into a useful device with practical applications.  The
effectiveness of theoretical input into this process has been
increasingly recognized~\cite{Fennie06p267602, Lee10p954}. First
principles prediction of the structure and properties of both real and
as-yet hypothetical materials can play a valuable role in the
screening of candidate materials for experimental investigation.  The
theoretical component of materials discovery and design can be further
strengthened by the effective use of crystallographic database
information to identify systems and structural families to be
screened. Our hope, by effective use of the database in conjunction
with first principles methods, is to enhance the value of our
theoretical input to experimental collaborators, who are the essential
contributors in any materials discovery/design effort, and thus to
expedite and increase the chances of success of the overall
process. However, effectively joining both database and first
principles methods to develop systematic and efficient strategies
requires experience through application to specific materials
challenges, such as the search for new ferroelectric materials.

Ferroelectric materials are a class of functional materials which are
particularly well suited to rational discovery and design approaches.
A ferroelectric is an insulating material characterized by a electric
polarization that is switchable by an applied electric field; this
polarization generally arises from a polar structural distortion of a
high-symmetry reference structure.  An applied electric field can thus
be used directly to manipulate the electric polarization of a
ferroelectric, and thus also control any properties, including strain,
magnetism, and optical response, that are coupled to the polarization.
Well-established technological applications of ferroelectrics include
nonvolatile information storage associated with the multiple
polarization states~\cite{Scott89p1400}, and transducers, exploiting
the piezoelectric response of the polar crystal structure and
polability of polycrystalline materials~\cite{Izyumskaya07p111}.

The optical and electronic properties of ferroelectrics offer further
possibilities for technological application, with potential impact on
electronics~\cite{Mannhart10p1608} and energy conversion and
storage~\cite{Clingman61p675}.  Of particular interest is the
possibility of electric field control of optical and electronic
properties via the switchable polarization. For example, in a doped
ferroelectric semiconductor, the polar distortion can couple to the
carriers and produce characteristic transport
properties~\cite{Edelstein95p2004} including a switchable diode effect
and a bulk photovoltaic effect, in which absorption of light by a
piezoelectric material generates an asymmetric carrier distribution
resulting in a net current~\cite{Fridkin01p654}.  These effects have
been recently observed \cite{Yang09p062909, Alexe11p256,
Katiyar11p092906, Choi09p63} in the ferroelectric perovskite oxide
BiFeO$_{3}$, the most-studied room-temperature multiferroic
\cite{Wang03p1719, Catalan09p2463}.

\begin{figure}

\centering
\includegraphics[width=3.0in]{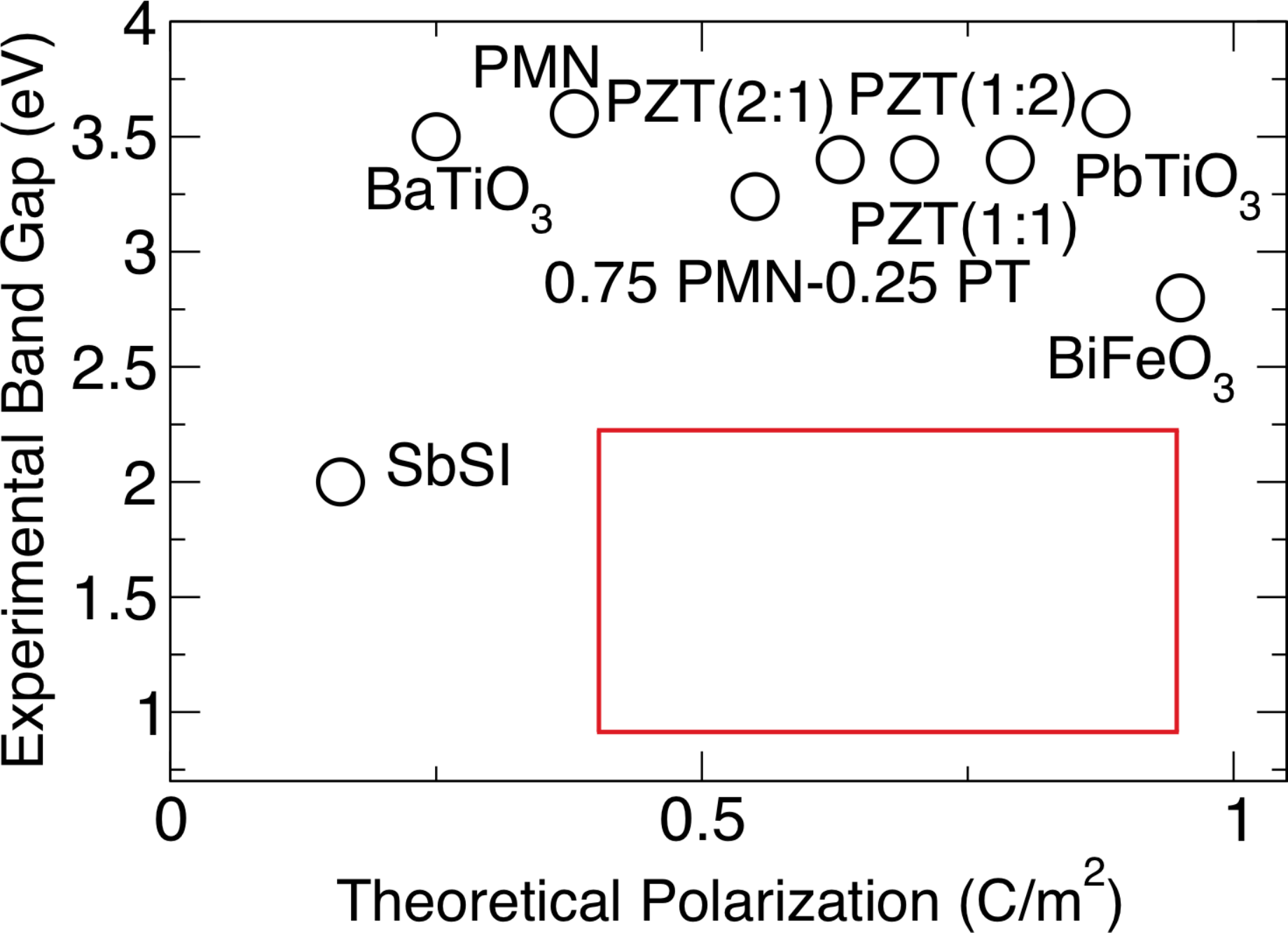}
\caption{The experimentally measured band gaps of a representative set
of ferroelectric materials are plotted against the value of
spontaneous polarization computed using first principles methods. The
empty box outlined in red highlights the lack of ferroelectric
materials with large polarization and smaller band gap comparable to
the solar spectrum. From Ref.~\cite{Bennett08p17409}.}
\label{fig:figure1}
\end{figure}

Ferroelectric semiconductors, with bandgaps in or below the visible
range, are therefore of particular interest.  BiFeO$_{3}$ has received
attention as it combines a high ferroelectric polarization with a
semiconductive band gap of 2.7 eV. However, as shown in
Figure~\ref{fig:figure1}, there are virtually no ferroelectric
materials with both a high polarization and a lower band gap, as
demonstrated by the empty space outlined by the red box. The challenge
thus is to discover materials in this region.

One route to increasing the number of ferroelectric semiconductors is
by band gap engineering: reducing the too-large gaps of known
ferroelectric materials through compositional substitution or
strain. Examples of the former include
Pb(Ti$_{1-x}$Ni$_{x}$)O$_{3-x}$~\cite{Bennett08p17409, Gou11p205115},
Ba(Ti$_{1-x-y}$Ce$_{x}$Pd$_{y}$)O$_{3-y}$~\cite{Bennett10p184106} and
PZN-type relaxors~\cite{Qi11p245206}. Recently it has been shown
\cite{Berger11p146804} that values of epitaxial strain that make
SrTiO$_3$ ferroelectric also significantly reduce its band gap from
the bulk value of 3.2 eV, greatly increasing its value as a
photocatalyst.

Another route to increasing the number of ferroelectric semiconductors
is by inducing ferroelectricity in nonpolar semiconducting compounds
with structures closely related to ferroelectrics. This can be done
through changes in composition or epitaxial strain that either
destabilize a polar mode, as in SrTiO$_{3}$~\cite{Haeni04p758} and
CaTiO$_{3}$~\cite{Eklund09p220101R} under epitaxial strain, or
destabilize one or more nonpolar modes that satisfy the symmetry
conditions to produce a polarization, resulting in an improper
ferroelectric material~\cite{Picozzi08p434208}. Examples of improper
ferroelectrics include YMnO$_{3}$, a layered structure in which a zone
boundary mode induces a polarization~\cite{Fennie05p100103R,
Kim10p092902}, and systems in which a polarization is induced by a
trilinear coupling of polarization to two oxygen octahedron rotation
modes: Ruddlesden-Popper phase
Ca$_3$(Ti,Mn)$_2$O$_7$~\cite{Benedek11p107204}, the double perovskite
(Na,La)(Mn,W)O$_{6}$~\cite{Fukushima11p12186}, and
(PbTiO$_{3}$)$_1$/(SrTiO$_{3}$)$_3$
superlattices~\cite{Bousquet08p732}.

While the two routes described above increase the number of
semiconducting ferroelectrics by incremental modifications of familiar
systems, there is also great interest in discovering novel
ferroelectric compounds. To do this, we extend the range of the search
beyond known ferroelectrics to all polar materials, because any polar
structure type also can in principle have representatives that are
ferroelectric. This requires that the energy of the polar distortion
which relates the polar subgroup structure to the nonpolar supergroup
structure be small in the sense that the corresponding energy barrier
can be overcome by electric fields less than the breakdown field of
the material.  Thus in the space of polar structures, we target
materials that have a low barrier to polarization switching as well as a
low band gap.

The design of new ferroelectric materials can therefore draw on design
principles for polar materials~\cite{Maggard03p27,
Marvel07p13963}. For example, one design principle that has been
considered is the combination of polar molecular units as counterions
that promote arrangements in which the polar units align to produce a
macroscopic polarization~\cite{Chang08p8511, Kim09p5335,
Chang09p6865}.

First, we focus on discovery of previously-unrecognized ferroelectrics
among systems reported in the crystallographic literature, since this
information is readily available in crystallographic databases such as
the Inorganic Crystal Structural Database (ICSD)~\cite{Belsky02p364}.
Analyses of polar oxides and sulfides have previously been carried
out~\cite{Atuchin04p411, Halasyamani98p2753}.  Our search for new
semiconducting ferroelectrics includes not only polar oxides and
sulfides, but the full range of polar systems in the ICSD. In addition
to a polar space group, we require a low energy difference from a
high-symmetry reference structure, to promote switchability, and a low
band gap.  In earlier work, Abrahams has searched for
previously-unrecognized ferroelectrics by combining the space group
criterion with a structural criterion as a proxy for
switchability\cite{Abrahams88p585, Abrahams96p790,
Abrahams06p26}. Here, we similarly search for proximity to a high
symmetry reference structure, and include the chemical requirement of
inclusion of a main group element to promote the lower band gap.  For
individual compounds of interest, we supplement the results from the
database with first principles calculations, obtaining not only
structural information, but the band structure and the energy barrier
to the high-symmmetry reference structure needed to identify
ferroelectric character in a system. Moreover, with recent advances in
the speed and accuracy of first-principles methods, it has become
practical to enhance and extend the information in the database by
first principles calculations of the structure and properties of both
real and hypothetical materials in a high-throughput
mode~\cite{Morgan05p296, Jain11p2295, Fischer06p641, Hautier10p3762,
Armiento11p014103, Trimarchi07p104113, Roy11preprint}, as exemplified
by recent searches for high-performance piezoelectrics
~\cite{Hautier10p3762, Armiento11p014103, Roy11preprint}.

In the past two years, we have been developing the integration of
database searches and first-principles methods to identify novel polar
compounds with desirable properties, with a particular focus on
semiconducting ferroelectrics. By attacking a series of pilot
problems, we have explored a variety of strategies to develop useful
search criteria and efficiently search for materials realizations.  In
this concept paper, after a survey of the polar space groups and major
polar structure types in the ICSD, we describe the results of three of
these exploratory investigations, and discuss the lessons learned that
have shaped our development of a systematic
method~\cite{Roy11preprint, Bennett12preprint_fe,
Bennett12preprint_opto}.

\section{\label{sec:level1} Methodology}

A wealth of information about naturally occurring crystal structures
and the relationship between structure and composition is inherent in
a comprehensive crystallographic database. Here, we use the
ICSD~\cite{Belsky02p364}, which contains 142,000 entries, with new
entries being added at the rate of about 7,000 each year. The entry
for a given system contains information about composition and the
lattice type and parameters; additional information includes space
group, occupied Wyckoff positions and structural parameters,
refinement data such as thermal factors, warnings and comments, and
structure type.  Online searches can select for the number and/or type
of distinct chemical elements in the compound, the space group, and a
variety of other fields. So, for example, one can list all ternary
compounds with a given space group, or all compounds with a given
stoichiometry.

As large as the major crystallographic databases are, they are still
incomplete in many ways. We use first principles total-energy and band
structure calculations, equally feasible both for real and
hypothetical structures, to complement the database information.
First, prediction of the lowest energy structure is a standard
application of first principles total-energy
methods~\cite{Yin82p5668}. For a given compound in the database, we
can compare the observed structure type with other likely candidate
structure types, minimizing the total energy with respect to the
structural parameters in each case. Comparison of the result for the
lowest energy structure and its structural parameters with the
experimental information in the database can confirm the latter or
alternatively cast it into question for further investigation.

In addition, for a given compound in the database we can compute
experimentally measurable physical properties that are not included,
such as band gap, magnetization, electric polarization, phonon
frequencies, dielectric constants, and piezoelectric
coefficients. Finally, we can search for additional representatives of
a structure type with desired properties by considering a large number
of candidate compositions and calculating the structure, stability and
relevant properties of the hypothetical compounds in a high-throughput
study.

First principles calculations can be performed with a variety of
widely available software packages.  For the calculations described in
this paper, we used
ABINIT~\cite{Gonze09p2582}. Optimized~\cite{Ramer99p12471}
norm-conserving~\cite{Rappe90p1227} pseudopotentials were generated
using the OPIUM code~\cite{Opium}. The k-grid for structural
optimizations was at least 4 $\times$ 4 $\times$
4~\cite{Monkhorst76p5188}. Additional details for each case will be
given below.

\section{\label{sec:level1} Results}

\subsection {\label{sec:level2} Survey of known polar systems}

We begin by surveying all polar systems in ICSD, extending the
previous surveys for polar oxide and sulfide
materials~\cite{Halasyamani98p2753, Atuchin04p411}. Of the 230 space
groups, 68 are polar. The number of entries reported in these 68 space
groups is 12,553, which is less than ten percent of the total number
of entries in the database (142,000).  Polar compounds can therefore
be said to be relatively rare.

These polar entries can be sorted into six crystal lattice systems,
each containing one or more crystal classes: hexagonal ($6mm$ and $6$,
with 2264 compounds representing 18.0$\%$ of the total), rhombohedral
($3mm$ and $3$, with 2411 compounds at 19.2$\%$), tetragonal ($4mm$
and $4$ with 851 compounds, 6.8$\%$), orthorhombic ($mm2$ with 4033
compounds, 32.1$\%$), monoclinic ($2$ and $m$ with 2534 compounds,
20.2$\%$) and triclinic ($1$ with 460 compounds, 3.7$\%$) entries.
The most familiar ferroelectric materials are rhombohedral
(LiNbO$_{3}$), hexagonal (YMnO$_{3}$) and tetragonal (PbTiO$_{3}$ and
BaTiO$_{3}$), yet there are twice as many polar orthorhombic entries
as there are either rhombohedral or hexagonal, and four times as many
polar orthorhombic entries than there are tetragonal entries.  This
result indicates that searches among orthorhombic compounds for
previously-unrecognized ferroelectrics might be particularly
rewarding, and indeed ferroelectricity might be favored by the lower
symmetry.
                          
\begin{table}
\begin{center}
\begin{tabular}{llllllllcl}
Crystal&Space&Total&1 &2 &3 &4 &5+ &Most Common Structure Types&Unassigned\\
Class &Group&Entries & & & & & & & \\
$6mm$ &186 &1327 &6 &374 &551 &248 &148 &LiGaGe (139), ZnS (445)&118 (8.9\%)\\
&185 &169 &0 &32 &67 &52 &18 &LuMnO$_{3}$ (54), LaF$_{3}$ (25)&46 (27.2\%)\\
&184 &15 &0 &0 &5 &2 &8 &zeolite (9)&5 (33.3\%)\\
&183 &19 &0 &1 &7 &9 &2 &none&19 (100\%)\\
&&&&&&&&&\\
$6$ &173 &681 &0 &29 &125 &304 &223 &Apatite (39), La$_{3}$CuSiS$_{7}$ (177)&145 (21.3\%)\\
&172 &1 &0 &0 &0 &1 &0 &none &1 (100\%)\\
&171 &5 &0 &3 &0 &1 &1 &none &5 (100\%)\\
&170 &13 &0 &1 &0 &7 &5 &Ba(NO$_{2}$)$_{2}$*2H$_{2}$O (5)&7 (53.9\%)\\
&169 &33 &0 &6 &9 &6 &12 &Al$_{2}$S$_{3}$ (11) &16 (48.5\%)\\
&168 &1 &0 &0 &0 &1 &0 &none &0 (0.0\%)\\
&&&&&&&&&\\
$3mm$ &161 &587 &0 &22 &212 &221 &132 &LiNbO$_{3}$ (287), whitlockite (61)&118 (20.1\%)\\
&160 &798 &0 &276 &215 &102 &205 &ZnS (123), FeBiO$_{3}$ (79)&216 (27.1\%)\\
&159 &168 &2 &27 &22 &57 &60 &Si$_{3}$N$_{4}$ (18) &80 (47.6\%)\\
&158 &22 &0 &5 &3 &7 &7 &none &22 (100\%)\\
&157 &72 &0 &3 &28 &26 &15 &Mg$_{3}$Si$_{2}$O$_{5}$(OH) (19)&30 (41.7\%)\\
&156 &347 &0 &259 &29 &53 &6 &CdI$_{2}$ (137)&116 (33.4\%)\\
&&&&&&&&&\\
$3$ &146 &214 &1 &12 &76 &76 &49 &Ni$_{3}$TeO$_{6}$ (12)&131 (61.2\%)\\
&145 &22 &0 &1 &7 &6 &8 &RbNO$_{3}$ (2)&19 (86.4\%)\\
&144 &82 &0 &16 &23 &32 &11 &RbNO$_{3}$ (16)&42 (51.2\%)\\
&143 &99 &1 &13 &33 &21 &31 &NiTi (6)&86 (86.9\%)\\
&&&&&&&&&\\
$4mm$ &110 &51 &0 &4 &28 &10 &9 &Li$_{2}$B$_{4}$O$_{7}$ (22) &11 (21.6\%)\\
&109 &50 &0 &11 &30 &6 &3 &LaPtSi (16) &19 (38.0\%)\\
&108 &29 &0 &1 &21 &5 &2 &Pb$_{5}$Cr$_{3}$F$_{19}$ (4) &25 (86.2\%)\\
&107 &121 &0 &13 &69 &17 &22 &BaNiSn$_{3}$ (43) &44 (36.4\%)\\
&106 &5 &0 &0 &2 &3 &0 &none &5 (100\%)\\
&105 &5 &0 &0 &4 &1 &0 &none &5 (100\%)\\
&104 &9 &0 &0 &5 &0 &4 &Tl$_{4}$HgI$_{6}$ (4) &5 (55.6\%)\\
&103 &8 &0 &6 &2 &0 &0 &NbTe$_{4}$ (7)&1 (12.5\%)\\
&102 &23 &4 &6 &5 &4 &4 &Al$_{2}$Gd$_{3}$ (4)&12 (52.2\%)\\
&101 &3 &0 &0 &3 &0 &0 &none &3 (100\%)\\
&100 &97 &0 &1 &16 &57 &23 &BSN (21), Ba$_{2}$TiOSi$_{2}$O$_{7}$ (23)&20 (20.6\%)\\
&99 &299 &0 &6 &104 &82 &107 &PbTiO$_{3}$ (210), PbVO$_{3}$ (12)&32 (10.7\%)\\
&&&&&&&&&\\
$4$ &80 &16 &0 &3 &6 &5 &2 &none &16 (100\%)\\
&79 &42 &0 &5 &20 &7 &10 &U$_{3}$Al$_{2}$Si$_{3}$ (6)&29 (69.1\%)\\
&78 &12 &1 &0 &6 &1 &4 &none &12 (100\%)\\
&77 &10 &0 &4 &4 &2 &0 &H$_{2}$S (3)&7 (70.0\%)\\
&76 &48 &2 &5 &22 &4 &15 &Ca$_{2}$P$_{2}$O$_{7}$ (16)&23 (47.9\%)\\
&75 &23 &0 &7 &4 &6 &6 &K$_{4}$CuV$_{5}$O$_{15}$Cl (4)&19 (82.6\%)\\
\end{tabular}
\caption{{ For each polar space group of hexagonal, rhombohedral and
tetragonal symmetry, arranged according to crystal class, we record
the total number of entries, the breakdown into entries containing 1,
2, 3, 4 or 5+ distinct chemical elements, the most common structure
types with the corresponding number of representatives, and the number
of entries for which no structure type is reported, also expressed as
a percentage of the total number of entries for the given space
group.}}
\label{table:polar1}
\end{center}
\end{table}
                          
\begin{table}
\begin{center}
\begin{tabular}{llllllllcl}
Crystal&Space&Total&1 &2 &3 &4 &5+ &Most Common Structure Types&Unassigned\\
Class &Group&Entries & & & & & & &\\
$mm2$ &46 &179 &0 &4 &52 &52 &71 &Ca$_{2}$AlFeO$_{5}$ (79), ErSr$_{2}$GaCu$_{2}$O$_{7}$ (27)&53 (29.6\%)\\
&45 &33 &0 &0 &15 &8 &10 &Ca$_{11}$InSb$_{9}$ (5), Sr$_{4}$Fe$_{6}$O$_{13}$ (15) &17 (51.5\%)\\
&44 &153 &0 &25 &78 &31 &19 &NaNO$_{2}$ (13)&74 (48.4\%)\\
&43 &287 &0 &43 &44 &87 &113 &Natrolite (61)&111 (38.7\%)\\
&42 &69 &1 &23 &21 &16 &8 &NbS$_{2}$ (10)&47 (68.1\%)\\
&41 &117 &0 &18 &24 &54 &21 &Bi$_{4}$Ti$_{3}$O$_{12}$ (15)&71 (60.7\%)\\
&40 &107 &0 &15 &16 &58 &18 &NaCu$_{2}$NbS$_{4}$ (14)&55 (51.4\%)\\
&39 &42 &0 &14 &15 &7 &6 &LaS (5)&31 (73.8\%)\\
&38 &186 &2 &51 &86 &37 &10 &CeNiC$_{2}$ (29)&108 (58.1\%)\\
&37 &22 &0 &0 &6 &4 &12 &none &22 (100\%)\\
&36 &691 &1 &65 &268 &214 &143 &Bi$_{3}$TiNbO$_{9}$ (87)&249 (36.0\%)\\
&35 &41 &0 &1 &6 &16 &18 &none &41 (100\%)\\
&34 &55 &0 &5 &8 &25 &17 &Ca$_{2}$B$_{5}$O$_{9}$Br (12) &35 (63.6\%)\\
&33 &1087 &2 &68 &300 &377 &340 &Cu$_{2}$Sc$_{2}$O$_{5}$ (18), NaFeO$_{2}$ (56)&258 (23.7\%)\\
&32 &48 &0 &5 &15 &10 &18 &K$_{1-x}$FeF$_{3}$ (4)&33 (68.8\%)\\
&31 &376 &0 &16 &136 &114 &110 &Cu$_{3}$AsS$_{4}$ (56)&157 (41.8\%)\\
&30 &20 &0 &2 &3 &5 &9 &Fe$_{3}$(PO$_{4}$)$_{2}$*H$_{2}$O &19 (95.0\%)\\
&29 &305 &1 &11 &83 &124 &86 &boracite (16)&184 (60.3\%)\\
&28 &30 &0 &6 &12 &2 &10 &AuTe$_{2}$ (5)&20 (66.7\%)\\
&27 &5 &0 &1 &3 &1 &0 &V$_{4}$H$_{3}$ (1)&4 (80.0\%)\\
&26 &133 &0 &9 &37 &59 &28 &NaNbO$_{3}$ (5)&88 (66.2\%)\\
&25 &47 &0 &17 &12 &12 &6 &GaAs (7)&29 (61.7\%)\\
&&&&&&&&&\\
$m$ &9 &621 &0 &40 &159 &215 &207 &Pb(Ti,Zr)O$_{3}$ (14) &339 (54.6\%)\\
&8 &350 &0 &30 &103 &116 &101 &Ca$_{5}$(BO$_{3}$)$_{3}$F (17), Pb$_{2}$FeNbO$_{6}$ (48) &199 (56.9\%)\\
&7 &307 &0 &51 &68 &108 &80 &WO$_{3}$ (32) &189 (61.6\%)\\
&6 &59 &0 &7 &13 &22 &17 &PMN-PT and (Na,K)NbO$_{3}$ (17) &42 (71.2\%)\\
&&&&&&&&&\\
$2$ &5 &430 &1 &30 &123 &152 &124 &many &292 (67.9\%)\\
&4 &722 &10 &35 &151 &282 &244 &many &415 (57.5\%)\\
&3 &45 &2 &4 &11 &13 &15 &none &45 (100\%)\\
&&&&&&&&&\\
$1$ &1 &460 &6 &52 &118 &150 &134 &many &324 (70.4\%)\\
\end{tabular}
\caption{{For each polar space group of orthorhombic, monoclinic, and
triclinic symmetry, arranged according to crystal class, we record the
total number of entries, the breakdown into entries containing 1, 2,
3, 4 or 5+ distinct chemical elements, the most common structure types
with the corresponding number of representatives, and the number of
entries for which no structure type is reported, also expressed as a
percentage of the total number of entries for the given space group.
The term ``many'' appearing under most common structure type means
that no one or two structure types have more representatives than a
number of others.}}
\label{table:polar2}
\end{center}
\end{table}

Next, we turn to closer examination of the composition and structure
of individual polar systems. Each entry can be characterized as
elemental (0.3$\%$), binary (14.3$\%$), ternary (29.9$\%$), quaternary
(30.4$\%$) or containing 5+ (25.1$\%$) distinct chemical elements;
this breakdown is given for each polar space group in
Tables~\ref{table:polar1}~and~\ref{table:polar2}. We divide the
breakdown of entries into two tables to emphasize the distinction
between crystal classes that contain common ferroelectrics
(Table~\ref{table:polar1}: hexagonal, rhombohedral, tetragonal) and
crystal classes in which few ferroelectrics are well known
(Table~\ref{table:polar2}: orthorhombic, monoclinic, triclinic).

It is intriguing to notice that there are reports in ICSD of elemental
systems with polar structures (the proportion 0.3$\%$, given above,
corresponds to 40 entries). On closer inspection, we find most of
these to be misassigned, with the given atomic positions giving a
higher-symmetry nonpolar structure within experimental error.
However, there are a few that appear to be truly polar, including Po
with a reported monoclinic $C2$(5) structure, and these warrant
further investigation.  Structures of systems with two or more
distinct chemical elements can be recognized either as a solid
solution which can be related to a simpler structure with fewer
distinct chemical elements, or as an inherently more complex
structure. In particular, many of the entries that have 5+ distinct
chemical elements have fractional occupation of Wyckoff positions,
indicative of a solid solution.

A more detailed classification of polar structures can be obtained by
grouping entries according to structure type.  Entries in ICSD report
the assignment from the original article (59.7$\%$ of entries), though
in many cases (40.3$\%$) none is made. In Tables~\ref{table:polar1}
and~\ref{table:polar2}, for each space group we list the most commonly
reported structure types and the number of representative entries for
each of these structure types. There are two space groups strongly
dominated by common structure types: hexagonal space group 186 and
tetragonal space group 99. In space group 186 the two most common
structure types are ZnS with 445 entries and LiGaGe with 139 entries,
which represent 44.0$\%$ of the 1327 total entries. In space group 99
the three most common structure types, PbTiO$_{3}$, (210 entries)
KNbO$_{3}$ (17 entries) and PbVO$_{3}$ (12 entries), are all related
to the perovskite structure and account for 79.3$\%$ of the 299
entries. At the other extreme, there are 13 space groups in which all
entries have no reported structure type: 183, 172, 171, 168, 158, 106, 105,
101, 80, 78, 37, 35 and 3; however the total number of entries in
these groups is only 196, or 1.6$\%$ of the total number of all polar
entries.

While a systematic classification system for the assignment of
inorganic structure type has been established~\cite{LimaDeFaria90p1},
it seems that these conventions are not always followed in the
literature, and thus not in the ICSD.  The case of tetragonal space
group 99 illustrates the problem of structure type nomenclature: the
similarity of the PbTiO$_{3}$ and KNbO$_{3}$ structures suggests that,
following the guidelines of Ref.~\cite{LimaDeFaria90p1}, these
should be combined and regarded as a single structure type. An
automated classification method based on the structural information in
ICSD would reduce or even eliminate such problems, and could also
assign missing structure types, especially in cases where the system
is chemically different from other representatives and the structural
relationship was not recognized by the authors of the original
paper. This would make the classification by structure type much more
useful, and in particular would make it possible to easily identify
unusual or exotic compounds with a given structure type of interest.

For the discovery of ferroelectric materials, these results suggest
two complementary strategies. The first is to search the database to
identify previously-overlooked systems in structure types of known
ferroelectrics and use first-principles methods to screen these
systems for exceptional properties. The second is to identify
additional structure types for which no ferroelectrics have been
previously reported, but for which ferroelectricity could be
established either in known representatives through further
experimental investigation, or in additional representatives
identified through a first-principles high-throughput study. We
develop both strategies in the three exploratory investigations
discussed next.
 
\subsection {\label{sec:level2} Sr(Sb$_{1/2}$Mn$_{1/2}$)O$_{3}$: A semiconducting ferroelectric?}
 
In our first exploratory investigation, we show how first principles
calculations can be used to complement a database search for
previously-overlooked ferroelectric oxides. In this case, the database
search was performed not by us, but previously by
Abrahams~\cite{Abrahams96p790} as part of a large-scale research
program to identify previously-overlooked ferroelectric materials in
the ICSD.  This search resulted in the identification of a number of
candidate ferroelecrics, including the double perovskite
Sr(Sb$_{1/2}$Mn$_{1/2}$)O$_{3}$~\cite{Foster97p3076} ; this system was
of particular interest to us because one-half of the octahedral units
contain a main group element (Sb) which could promote a lower band
gap, making the system a ferroelectric semiconductor. In addition, the
presence of the transition metal Mn could lead to magnetic and
multiferroic behavior. However, since this search was performed in
1996, there have been at least five additional reports of the same
system in various, mostly nonpolar, tetragonal space groups, calling
the ferroelectric character of the system into question. This suggests
that it would be illuminating to study the structural energetics of
this system from first principles.

The original report~\cite{Politova91p2017} assigns room temperature
Sr(Sb$_{1/2}$Mn$_{1/2}$)O$_{3}$ to polar space group $I4mm$ (107) with
full rocksalt ordering for the Sb and Mn on the B site. The ICSD
includes several later entries for this system, reporting different
tetragonal space groups including nonpolar $I4/mcm$ (140, no cation
order)~\cite{Lufaso04p1651}, and nonpolar $I4/m$ (87, partial rocksalt
cation ordering)~\cite{Cheah06p1775, Mandal08p2325, Ivanov09p822}. A
value for the band gap of 0.50(2)~eV was extracted from the Arrhenius
behavior of the conductivity in Ref.~\cite{Foster97p3076} (in which
the space group was reported to be polar $I4mm$, consistent with
Ref.~\cite{Politova91p2017}).

\begin{figure}
\centering
\includegraphics[width=3.0in]{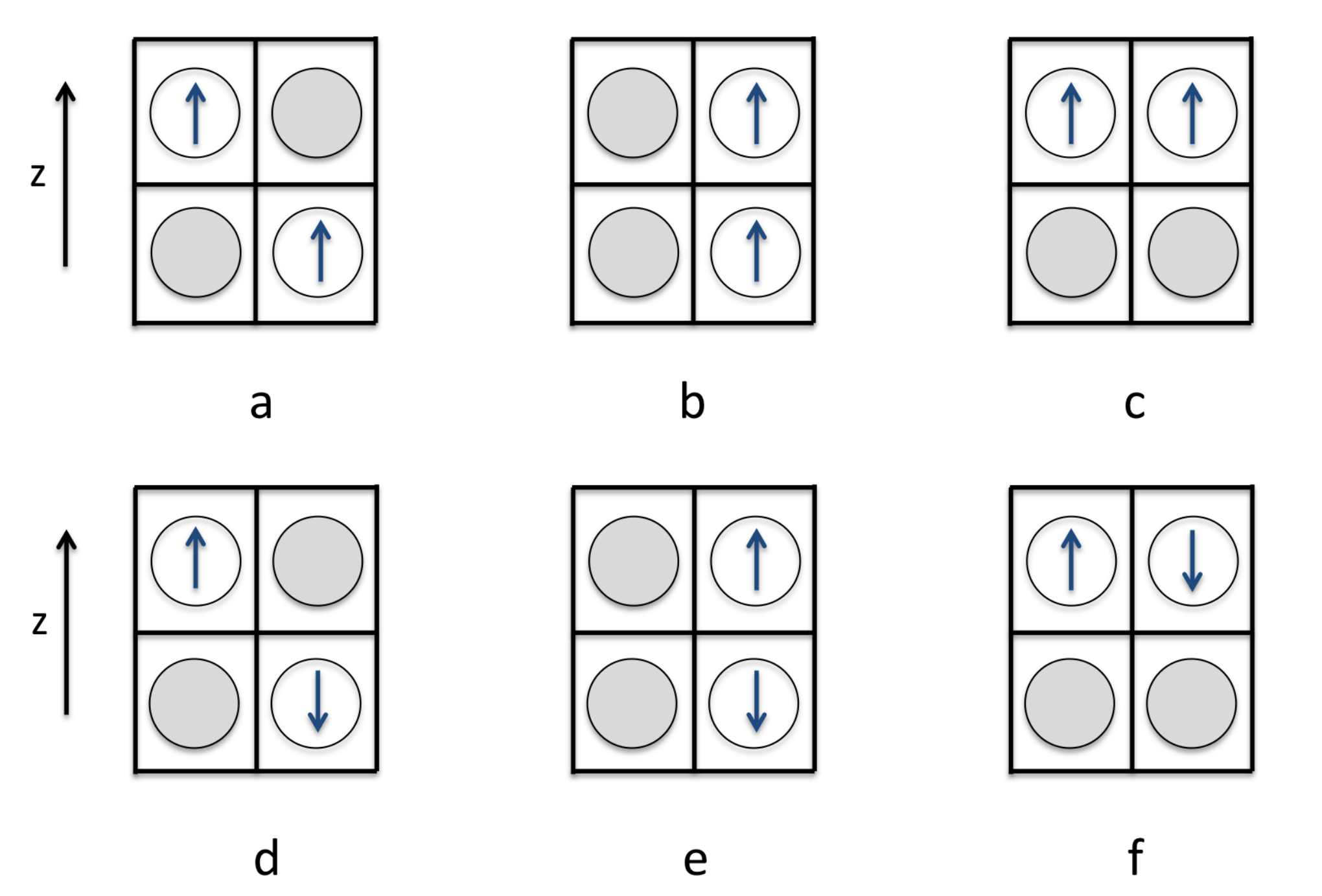}
\caption{Sketches of the magnetic and cation ordering for the six
Sr(Sb$_{1/2}$Mn$_{1/2}$)O$_{3}$ configurations considered. a through c
are FM and d through f are AFM ordered rocksalt, pillars and layers,
respectively. The boxes represent perovskite unit cells, Sb are shown
as shaded circles, and the spin state of each Mn (open circles) is
shown with an arrow as spin up or down.}
\label{fig:figure2}
\end{figure}

To clarify the question of space group assignment and polar character
of Sr(Sb$_{1/2}$Mn$_{1/2}$)O$_{3}$, we begin by computing total
energies and optimized structural parameters for various choices of
magnetic and cation ordering. We consider three types of cation
ordering (rocksalt, 2D checkerboard of Mn and Sb pillars, and
alternating layers of Mn and Sb) and two types of magnetic ordering on
the Mn sublattice, ferromagnetic (FM) and antiferromagnetic (AFM), as
shown in Figure~\ref{fig:figure2}. All six combinations can be
accommodated by a $\sqrt{2} \times \sqrt{2} \times$ 2 supercell. The
total energy calculations and structural relaxations are performed
with a starting structure with atomic positions and cell parameters
taken from Ref.~\cite{Politova91p2017}, as given in Table 1 of
Ref.~\cite{Foster97p3076}, combined with an additional oxygen
octahedron rotation around $z$, alternating from plane to plane, taken
from Ref.~\cite{Lufaso04p1651}.  As will be further discussed below,
this choice guarantees that the structure can relax to any of the
experimentally observed structure types.  Examination of the relaxed
structure obtained from this starting point will reveal if any
symmetries are broken beyond those determined by the cation and
magnetic ordering.

\begin{table}
\begin{center}
\begin{tabular}{llllllllllll}
& $a$ & $c$   &$\Delta E$& $\Gamma_{4}^{-}$ & $R_{5}^{+}$ & $X_{3}^{-}$ & $M_{2}^{+}$ & $M_{4}^{+}$ & $R_{2}^{-}$ & $R_{3}^{-}$ & $R_{5}^{-}$\\
Expt. ($I4mm$)& 5.526 &8.039&-&0.346 &0.020 &0 &0 &0 &0.160 &0.054  &0\\
Expt. ($I4/m$)& 5.533 &8.085&-&0 &0 &0 &0 &0 &0.003 &0.093 &0.385\\
Expt ($I4/mcm$)& 5.556 &8.055&-&0 &0 &0 &0 &0 &0 &0 &0.332\\
rocksalt          &           &&         &        &        &        &        &        & & &\\
FM ($I4/m$)   & 5.548 &8.135&0.12&0 &0 &0 &0 &0 &0.095 &0.029 &0.614\\
AFM ($I4/m$)  & 5.459 &8.299&0.13&0 &0 &0 &0 &0 &0.108 &0.048 &0.702\\
pillars     &           &&         &        &        &        &        &   & & &\\
FM ($P4/mcc$) & 5.536 &8.212&0.41&0 &0 &0 &0 &0.066 &0 &0 &0.630\\
AFM ($P4/m$)  & 5.539 &8.133&0.50&0 &0 &0.023 &0.005 &0.083 &0.022 &0.022 &0.636\\
layers  &           &&         &        &        &        &        &   & & &\\
FM ($P4/mbm$) & 5.459 &8.208&0&0 &0 &0.274 &0.426 &0 &0 &0 &0.492\\
AFM ($P4/mbm$)& 5.520 &8.190&0.08&0 &0 &0.297 &0.442 &0 &0 &0 &0.520\\
\end{tabular}
\caption{Structural data for Sr(Mn$_{1/2}$Sb$_{1/2}$)O$_{3}$ from
experimental reports and from first principles calculations for the
six ordered configurations considered. Lattice constants $a$ and $c$
are in \AA. $\Delta E$ is the computed energy per 10-atom formula
unit, in eV, relative to the minimum energy FM layered
configuration. The remaining columns give the amplitude of each normal
mode distortion needed to specify the atomic positions, as discussed
in the text.}
\label{table:SSM}
\end{center}
\end{table}

To describe the relaxed structures in a way that allows us directly to
compare all six magnetic and cation orderings, we use the amplitudes
of symmetrized displacement patterns defined with respect to the
high-symmetry reference structure where the atoms are at the ideal
cubic perovskite positions and all $B$ site cations are treated as
symmetry equivalent. For a given relaxed structure, we obtain the
amplitudes of the contributing patterns using the program ISODISTORT,
which establishes the conventions for the perovskite
modes~\cite{Campbell06p607}. 

The cation and magnetic orderings of the ordered supercells break
symmetries that induce atomic displacements of corresponding symmetry
types, as follows: for the rocksalt ordering, the breathing pattern
$R_{2}^{-}$, in which the surrounding oxygens move in towards one $B$
cation and out from the other; for pillar ordering, the $xy$-plane
breathing pattern $M_{4}^{+}$, in which the surrounding equatorial
oxygens move in towards one $B$ cation pillar and out from the other;
and for the alternating layers, the $z$-breathing pattern $X_{3}^{-}$,
in which apical oxygens move towards one $B$ cation layer and away
from the other. Note that in the absence of spin-orbit coupling, the
crystallographic symmetry of the FM and AFM structures is the same.

The symmetrized displacement patterns that comprise the starting
structure described earlier are $\Gamma_{4}^{-}$ (a polar mode with
displacements along $z$), the breathing mode $R_{2}^{-}$, the
Jahn-Teller mode $R_{3}^{-}$ (where equatorial oxygens move in towards
and apical oxygens move away from one $B$ cation) and the oxygen
octahedron rotation mode $R_{5}^{-}$, with rotations around $z$
alternating from plane to plane.  For the layer and pillar supercells,
cation ordering breaks additional symmetries as described above. In
the course of relaxation, the distortion patterns that do not lower
the energy will relax to zero, restoring the corresponding symmetry.

The results are presented in Table~\ref{table:SSM}.  It is immediately
clear that the oxygen-octahedron rotation $R_{5}^{-}$ is a strong
instability independent of cation and magnetic ordering, while the
ferroelectric distortion does not appear in any of the ordered
supercells.  To be more specific, for rocksalt ordering, both FM and
AFM systems relax to a tetragonal structure which combines the
$R_{2}^{-}$ breathing mode driven by the cation ordering with oxygen
octahedron rotations corresponding to the $R_{5}^{-}$ mode, which
appears to be the driving instability, and a small amplitude for the
$R_{3}^{-}$ Jahn-Teller distortion that then appears without
additional symmetry breaking; the $\Gamma_{4}^{-}$ mode introduced by
the starting positions relaxes to zero. This structure has the cation
ordered $I4/m$ space group (87) identified in several experimental
investigations~\cite{Mandal08p2325, Ivanov09p822}. Similarly, for
layered cation ordering, both systems relax to a tetragonal structure
which combines the $X_{3}^{-}$ breathing mode driven by the cation
ordering with oxygen octahedron rotations generated by a combination
of the $R_{5}^{-}$ and $M_{2}^{+}$ modes, which produce rotations by
different angles in the Sb and Mn layers. Finally, for the pillar
ordering, the FM system relaxes to a tetragonal structure which
combines the $M_{4}^{+}$ breathing mode driven by the cation ordering
with oxygen octahedron rotations corresponding to the $R_{5}^{-}$
mode. However, the AFM system has a small symmetry-breaking
instability relative to the higher-symmetry structure imposed by the
cation ordering and octahedron rotation mode alone, as confirmed by
comparison of the total energy of the given structure with that of the
optimized higher-symmetry structure. This is reminiscent of the
spin-phonon coupling previously studied in other perovskite oxides,
including EuTiO$_3$~\cite{Fennie06p267602} and
SrMnO$_{3}$~\cite{Lee10p207204}, in which a change in magnetic order
can destabilize a phonon and lead to a symmetry-lowering distortion.

Our calculated ground state structure is the FM ordered alternating
Sb$^{5+}$/Mn$^{3+}$ layered configuration. This ordering differs from
that observed experimentally (rocksalt or partial rocksalt), and is
further surprising in that it appears electrostatically
unfavorable. However, its energy could be low due to the size mismatch
between the two B-sites: (Sb$^{5+}$ is 0.60~$\AA$ and Mn$^{3+}$ is
0.65~$\AA$) and correspondingly large relaxations, and it should be
noted that the observed ordering, when present, reflects the synthetic
conditions at high temperature rather than the ground state. Both
rocksalt structures are roughly 0.12~eV per 5-atom formula unit higher
in energy than the ground state, with the energy difference between FM
and AFM very small (5~meV), while the pillar structures are much
higher in energy. It is difficult to compare the predicted magnetic
ordering with experiments; the reported ordering is a spin glass
magnetic order and likely is related at least in part to the cation
disorder.

For the six configurations, we computed the electronic band structure
and found that all are metallic. When the band structure is recomputed
with LDA+$U$ with values of $U >$ 3~eV, a band gap is opened in the
minority-spin channel, but not in the majority-spin channel, yielding
a spin-polarized half-metal.

From this analysis, we conclude that Sr(Sb$_{1/2}$Mn$_{1/2}$)O$_{3}$
is, in fact, neither semiconducting nor ferroelectric as proposed in
Ref.~\cite{Foster97p3076}, but has a nonpolar structure, as reported
in subsequent papers for both disordered and partially ordered
cases. This negative result, though disappointing when considered in
the framework of the discovery of new ferroelectric materials, does
illustrate the value of including first principles results in
evaluating structural data.

\subsection {\label{sec:level2} Schafarzikite: $M$Sb$_{2}$O$_{4}$ ($M$=Mn, Fe, Ni, Zn)}

In our second exploratory investigation, we consider the
schafarzikites, a family of complex oxides reported with a nonpolar
structure. We choose this structure because it contains magnetic
elements within oxygen octahedra, that, unlike the conventional
perovskite, are edge-sharing. In addition, the Sb are not within the
octahedra, as in the previous example, but part of a chain that
cross-links the octahedra. In this way, we have two cations of
interest in two chemically different environments. We use first
principles results to investigate whether the ground state structure
of one or more compounds in this family might in fact be in a
lower-symmetry polar space group.  This could be the case if the
reported nonpolar structure is either the result of a
temperature-driven phase transition or simply a misassignment to a
higher-symmetry space group. The latter possibility was pointed out to
us by Prof. J. F. Scott of the University of Cambridge~\cite{JFScott},
based on the crystallographic practice of assigning a structure to the
highest-symmetry space group consistent with the structural data. This
suggests that some, perhaps many, polar compounds might be misreported
in the database with higher (nonpolar) space groups.

The question of whether known nonpolar structure families could have
representatives with polar instabilities has proved to be perennially
fascinating, especially with regard to spinels and
pyrochlores~\cite{Schmid74p2697, Fennie07preprint, Rabe10p211}. To
choose a distinct nonpolar structure type for the present study, we
searched ICSD for ternary antimony oxides, the main group element Sb
being, as in the previous section, selected to promote a lower band
gap in the hope of discovering new ferroelectric semiconductors. From
these, we chose to investigate schafarzikites, whose edge-sharing
oxygen octahedron chains aligned along a single direction might allow
a polar distortion along the chain.

\begin{figure}
\centering
\includegraphics[width=3.0in]{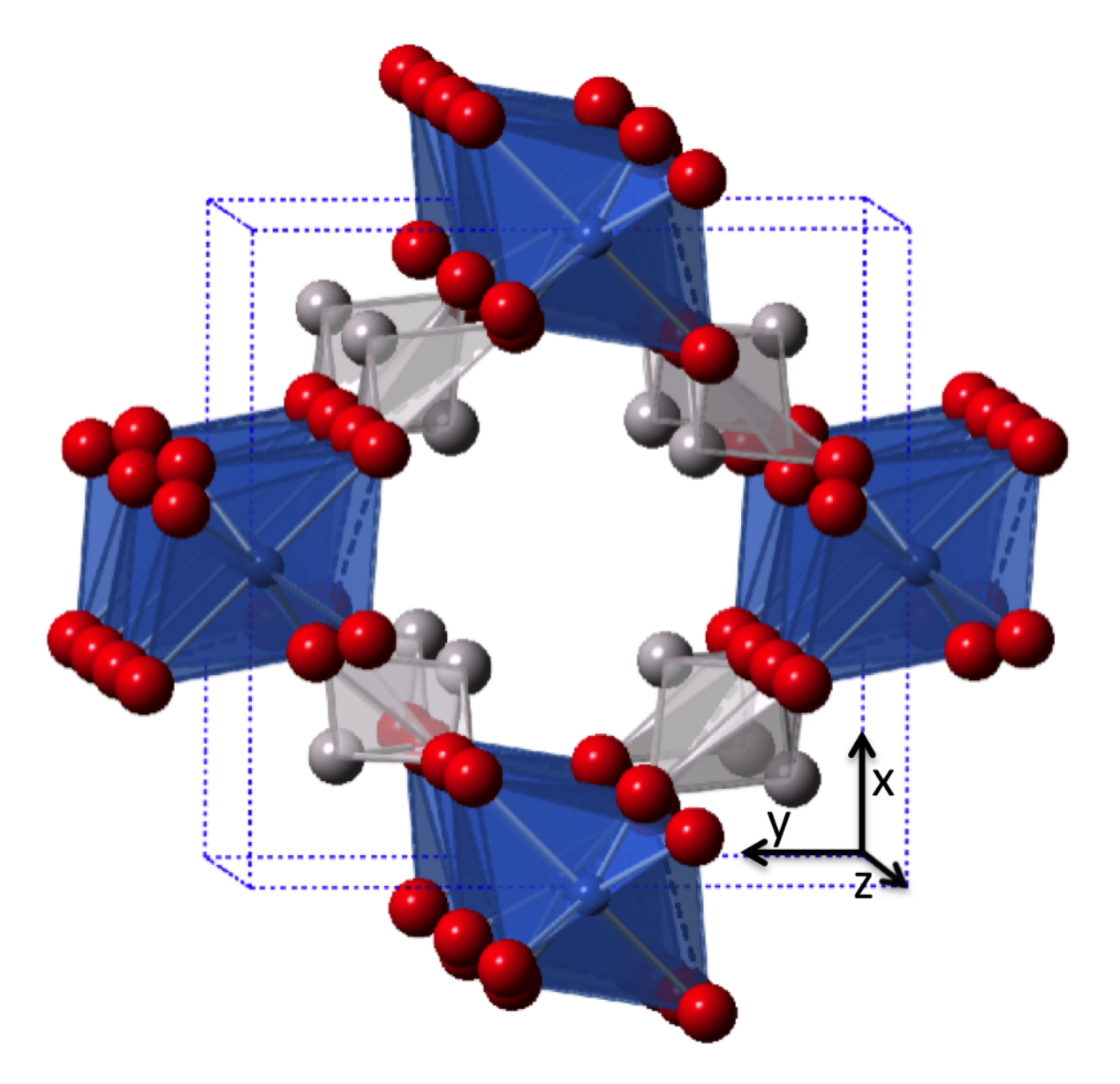}
\caption{The MSb$_{2}$O$_{4}$ schafarzikite structure contains chains
of corner-sharing oxygen octahedra (blue) that are cross-linked by
SbO$_{3}$ units (gray).}
\label{fig:figure3}
\end{figure}

The structure of the schafarzikite family $M$Sb$_{2}$O$_{4}$ is shown
in Figure~\ref{fig:figure3}. It has chains of edge-sharing $M$-centered
octahedra, crosslinked by chains of Sb ions in trigonal pyramidal
coordination with three oxygen atoms from the octahedral chains and
its own lone pair. For $M$=Fe, $A$-type antiferromagnetism
(anti-aligned along chains, aligned within the $ab$ plane) is observed
with $T_{N}$ = 45 K. An interesting possibility has been
raised~\cite{Whitaker11p14523} that doping on the Sb site could
functionalize the material by controlling the electronic state of the
transition metal cation, and hence its electrical and magnetic
properties. This, in turn, could couple to polar instabilities,
producing a multifunctional material.

Known representatives of the schafarzikite family have tetragonal
space group 135 ($P4_{2}/mbc$), in the Pb$_{3}$O$_{4}$ structure
type. ICSD contains entries in this structure type for compounds that
contain Mn, Fe, Ni, and Zn, all of which are formally 2$^{+}$
cations. Zn$^{2+}$ is not magnetic, but Mn, Fe and Ni should have
magnetic interactions. For these compounds, we study FM and three
types of AFM order: AFM alignment along the $c$ chains and FM
interchain in the $ab$ plane ($A$ type, magnetic group 28:
$4^{\prime}/m^{\prime}m^{\prime}m^{\prime}$), FM alignment along the
$c$ chains and AFM interchain in the $ab$ plane ($C$ type, magnetic
group 27: $4^{\prime}/mmm^{\prime}$), and AFM alignment along the $c$
chains and AFM interchain in the $ab$ plane ($G$ type, magnetic group
25: $4/m^{\prime}m^{\prime}m^{\prime}$.  Results of the structural
relaxations are shown in Table~\ref{table:schafarzikite}.

\begin{table}
\begin{center}
\begin{tabular}{llllll}
   & $a$   & $c$   & $E_{\rm gap}$ & $\mu$ & $\Delta$E\\ 
Mn & & &&&\\
FM & 8.533 (-1.9\%)& 6.125 (+2.4\%)&0.27&4.98& 0.251\\ 
$A$  & 8.546 (-1.7\%)& 6.050 (+1.1\%)&0.45&0& 0\\
$C$  & 8.531 (-1.9\%)& 6.099 (+2.0\%)&0.29&0& 0.101\\
$G$  & 8.568 (-1.5\%)& 6.044 (+1.0\%)&0.27&0& 0.114\\
Fe & & &&&\\
FM & 8.273 (-3.7\%)& 6.015 (+1.7\%)&0&4.62& 0.054\\
$A$  & 8.281 (-3.6\%)& 6.021 (+1.8\%)&0&0& 0.265\\
$C$  & 8.268 (-3.7\%)& 6.011 (+1.7\%)&0&0& 0\\
$G$  & 8.288 (-3.5\%)& 6.027 (+1.9\%)&0&0& 0.264\\
Ni & & &&&\\
FM & 8.293 (-0.9\%)& 6.031 (+2.1\%)&0.23&4.08& 0.106\\
$A$  & 8.288 (-1.0\%)& 6.027 (+2.0\%)&1.05&0& 0.015\\
$C$  & 8.280 (-1.1\%)& 6.021 (+1.9\%)&0.93&0& 0\\
$G$  & 8.293 (-0.9\%)& 6.030 (+2.1\%)&0.23&0& 0.154\\
Zn & & &&&\\
- & 8.487 (-0.5\%)& 5.873 (-1.16\%) &1.92&0& 0\\ 
\end{tabular}
\caption{First-principles lattice constants for the four schafarzikite
compounds $M$Sb$_2$O$_4$ ($M$ = Mn, Fe, Ni, Zn) are reported in \AA,
with error relative to experimental value in parentheses. The
calculated band gap is given in eV. For the magnetic compounds ($M$ =
Mn, Fe, Ni), results are given for four different magnetic orderings:
ferromagnetic (FM) and three different antiferromagnetic orderings
described in the text ($A$ type, $C$ type and $G$ type); for the FM
configuration, the magnetic moment per $M$ is given in $\mu_B$.  }
\label{table:schafarzikite}
\end{center}
\end{table}

From Table~\ref{table:schafarzikite}, it can be seen that the first
principles results for the magnetic representatives are not in good
agreement with available experimental data.  The $c/a$ ratio is
overestimated and for $M$=Fe, our DFT results show zero band gap and a
different magnetic ground state than that reported in
Ref.~\cite{Whitaker11p14523}. This suggests that at the least, a
DFT+$U$ calculation would be needed for an accurate description, and
highlights the importance of choosing an appropriate functional in
high-throughput first-principles studies. However, for non-magnetic
ZnSb$_{2}$O$_{4}$ we find good agreement with experimental lattice
constants and structural parameters~\cite{Puebla82p2020}.

Now, we investigate whether the ground state structure of
ZnSb$_{2}$O$_{4}$ might in fact be in a lower-symmetry polar space
group. We consider the polar structures which are related to the
high-symmetry nonpolar $P4_{2}/mbc$ structure by freezing in a single
normal mode; these are the most relevant in the search for new
ferroelectrics. The modes that generate polar structures in this way
can be identified through symmetry analysis, for which we use the
software package ISOTROPY~\cite{ISOTROPY}. More specifically, we list
the normal modes at the high-symmetry wavevectors $\Gamma$(0,0,0),
$X$({$\pi \over a$},0,0) and $Z$({0,0,$\pi \over c$}). We consider the
structure obtained by freezing in a single given mode, find the
resulting space group and check to see if it is one of the 68 polar
groups. First principles total-energy calculations are then performed
to relax each of these polar structures to determine whether the polar
distortion lowers the energy. This approach is much more
computationally efficient than the alternative of calculating the full
phonon dispersion and checking the symmetry character of the unstable
branches, because the symmetry analysis targets only the modes of
interest.

Application of this symmetry analysis to the schafarzikite structure
yields the following list of candidate polar structures. Zone center
modes $\Gamma_{3}^{-}$ and $\Gamma_{5}^{-}$ decrease symmetry to
$P4_{2}bc$ (106) and $Ama2$ (40), $Pmc2_{1}$ (26) or $Pm$ (6)
respectively; zone boundary modes $Z1$, $Z2$, $X1$ and $X2$ decrease
symmetry to $Pba2$ (32), $Pnn2$ (34), $Pca2_{1}$ (29) (or $Pc$ (7)),
and $Pmc2_{1}$ (26) (or $Pm$ (6)) respectively.

Using first principles calculations, we investigated the stability of
the polar $P4_{2}bc$ (106) structure, obtained by a zone center
$\Gamma_{3}^{-}$ mode, which induces polarization along $c$.  We froze
in a $\Gamma_{3}^{-}$ distortion, breaking the symmetry to $P4_{2}bc$,
and performed a structural relaxation.  For $M$=Zn, we found that
symmetry breaking distortions to this polar space group relax back to
the original non-polar structure, confirming the correct assignment of
the nonpolar space group in this case. Analogous calculations for our
ground state structures for $M$=Fe and Ni yielded the same result,
though as noted above, calculations with DFT+$U$ would be necessary
for a definitive result for these two systems.

The zone boundary modes that produce polar space groups in this
structure are especially interesting, as an instability to one of
these modes would make the system an improper ferroelectric. It is
extremely unlikely, though, that this would yield a bulk structure for
the compounds considered, as the cell doubling would be readily
manifest in any structural determination. However, it might be
possible to induce an instability either in the zone center polar
modes or in one of the zone boundary modes by perturbations such as
compositional substitution or epitaxial strain, and we believe this
could reward further investigation.

The fact that it turns out that the particular system chosen for this
example was in all likelihood correctly assigned as nonpolar in the
original literature, amounting to a negative result in the present
search for new ferroelectrics, does not invalidate the initial
hypothesis that some, perhaps many, polar compounds have been
misreported as nonpolar. Systematic consideration of a number of
nonpolar structure types would be needed to establish how prevalent
such misassignment or low-temperature phase transitions might be, and
to develop principles to identify nonpolar structure types with a
tendency to polar instabilities. However, the identification of routes
to improper ferroelectricity in this relatively low-symmetry nonpolar
structure is intriguing, and further investigation in other
low-symmetry nonpolar structure families seems warranted.

\subsection {\label{sec:level2} $M_{2}$P$_{2}$(S,Se)$_{6}$ chalcogenides}

In our final exploratory investigation, we use a known semiconducting
ferroelectric chalcogenide as a starting point for a database search
which identifies both additional representatives in the family that
could show a ferroelectric transition and polarization, and a
different family of compounds with the same stoichiometry that also
shows previously reported, though largely overlooked, indications of
ferroelectricity.

Materials which contain chalcogenide anions, such as S, Se and Te,
have a lower band gap than their O analogues, because the larger, more
polarizable chalcogenide anions form bonds to metals that are more
covalent and less ionic in nature than O. This principle has been of
great interest in the field of solar energy harvesting, as materials
based on the chalcopyrite structure (CuFeS$_{2}$) such as
Cu(In$_{1-x}$Ga$_{x}$)Se$_{2}$ (CIGS), can be readily processed as
films and nanostructures. This principle has also stimulated interest
in perovskite sulfides such as BaZrS$_{3}$, which has a theoretical
bandgap of 1.7~eV, lower than that calculated for the corresponding
oxide BaZrO$_{3}$ (3.9~eV)~\cite{Bennett09p235115}. However, the
chalcopyrites and known perovskite sulfides have nonpolar structures.

\begin{figure}
\centering
\includegraphics[width=3.0in]{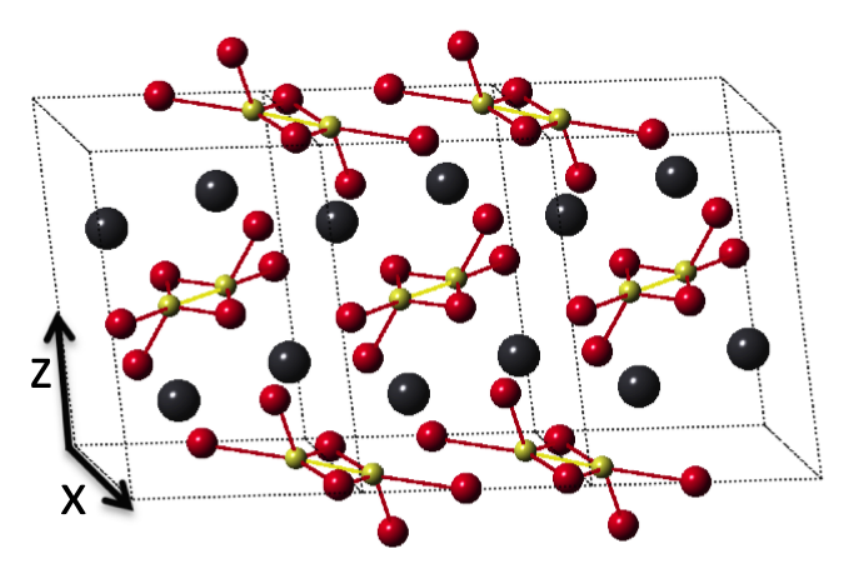}
\caption{The structure of Pb$_{2}$P$_{2}$Se$_{6}$ in monoclinic $Pc$
symmetry has alternating layers of Pb ions (black) and
P$_{2}$Se$_{6}$. Each P (gold) is bound to four Se (red) in a
distorted tetrahedral arrangement. Polarization arises from asymmetric
displacements of Pb along the (101) direction.}
\label{fig:figure4}
\end{figure}

A paraelectric-ferroelectric transition has been identified in the
semiconducting chalcogenides $M_{2}$P$_{2}X_{6}$ ($M$=Sn, Pb and
$X$=S, Se), and well studied by a variety of methods
~\cite{Moriya98p3505, vanLoosdrecht93p6014, Eijt98p4811,
Maior94p11211, Smirnov00p15051, Rushchanskii07p207601}. Further
interest was stimulated by the observation of a photovoltaic effect in
Sn$_{2}$P$_{2}$S$_{6}$ crystals~\cite{Cho01p3317}. First principles
studies of Sn$_{2}$P$_{2}$Se$_{6}$ show $E_{\rm gap}$ $\approx$ 1~eV,
and P is 0.15 C/m$^{2}$~\cite{Caracas02p104106}.  The structure of
these compounds is shown in Figure~\ref{fig:figure4}. Layers of $M$
ions alternate with layers of P$_{2}$X$_{6}$ units, in which each P is
tetrahedrally bound to four chalcogenide anions.  The polarization
arises from asymmetric displacements of the $M$-sites (Sn$^{2+}$ or
Pb$^{2+}$), both of which contain stereochemically active lone
pairs. The displacement of the $M$-sites causes a decrease in symmetry
from nonpolar space group $P2_{1}/c$ (14) to polar space group $Pc$
(7).

To identify other members of the $M_{2}$P$_{2}X_{6}$ family, we
searched ICSD for 3-element compounds MP(S,Se) that have 1:1:3
stoichiometry and include P and either S or Se.  Most of the entries
have nonpolar structures. In space group $P2_{1}/c$ (14) there are six
selenides (M=Sn, Pb, Ca, Sr, Ba and Eu) and one sulfide,
Sn$_{2}$P$_{2}$S$_{6}$, which all have the paraelectric structure
discussed above.  In space group $P3_{1}21$ (152) there are two
selenides (M=K,Rb), in $R{\bar3}$ (148) there are four selenides
(M=Mg, Zn, Fe, Mn), in $Immm$ (71) there are four sulfides (M=K, Rb,
Cs, and Tl), and in $C2/m$ (12) there are seven sulfides (M= Mg, Zn,
Cd, Ag, Fe, Mn, Ni). The structures of the mercury compounds are
unique: $C2/c$ (15) for Hg$_{2}$P$_{2}$Se$_{6}$ and $P{\bar1}$ (1) for
Hg$_{2}$P$_{2}$S$_{6}$.

In contrast, the number of entries with polar structures is much
lower: in $Pc$ (7) there are the four ferroelectric compounds,
Sn$_{2}$P$_{2}$(S,Se)$_{6}$, Pb$_{2}$P$_{2}$(S,Se)$_{6}$, discussed
above. K$_{2}$P$_{2}$Se$_{6}$ is reported in space group $P3_{1}$
(144), and Cd$_{2}$P$_{2}$S$_{6}$ and Fe$_{2}$P$_{2}$Se$_{6}$ in group
$R3$ (146).

This list suggests three different avenues for investigating polar
compounds and possible ferroelectricity in these systems. The first is
to use first-principles calculations to investigate whether the
selenides with M = Ca, Sr, Ba and Eu might have ground state
ferroelectric structures analogous to those of $M$ = Sn and Pb. This
is not very promising, as the proposal that a lone pair is necessary
to generate the asymmetric displacements needed for the
polarization~\cite{Glukhovpreprint, Caracas02p104106} seems quite
reasonable. Our list also contains two additional structure types that
show indications of ferroelectricity. Thus, another avenue would be
further to investigate the polar compound
K$_{2}$P$_{2}$Se$_{6}$. Unfortunately, this structure is a polymeric
helical crystal with over one hundred atoms per unit cell, and
therefore is not well-suited for full first-principles investigation.

The third avenue is the most promising: to investigate the
rhombohedral compounds that show indications of ferroelectricity, as
the two $R3$ compounds Cd$_{2}$P$_{2}$S$_{6}$ and
Fe$_{2}$P$_{2}$Se$_{6}$ have relatively simple structures. ISOTROPY
analysis shows that space group 146 is related to the higher symmetry
group 148 by a zone-center polar $\Gamma_{1}^{-}$ mode.  In the
nonpolar $R{\bar3}$ (148) space group, in addition to a candidate
paraelectric partner reported for the Fe compound, there are three
additional compounds already listed above: the selenides with $M$=Mg,
Zn, and Mn.  These nonpolar phases have been well studied
~\cite{Boucher95p952, Ouvrard85p1181}. Their structures are related to
the CdI$_{2}$ structure type, which can be described as layers of
edge-sharing Cd-centered halogen octahedra stacked in an $AB$ order
sequence.

\begin{figure}
\centering
\includegraphics[width=2.0in]{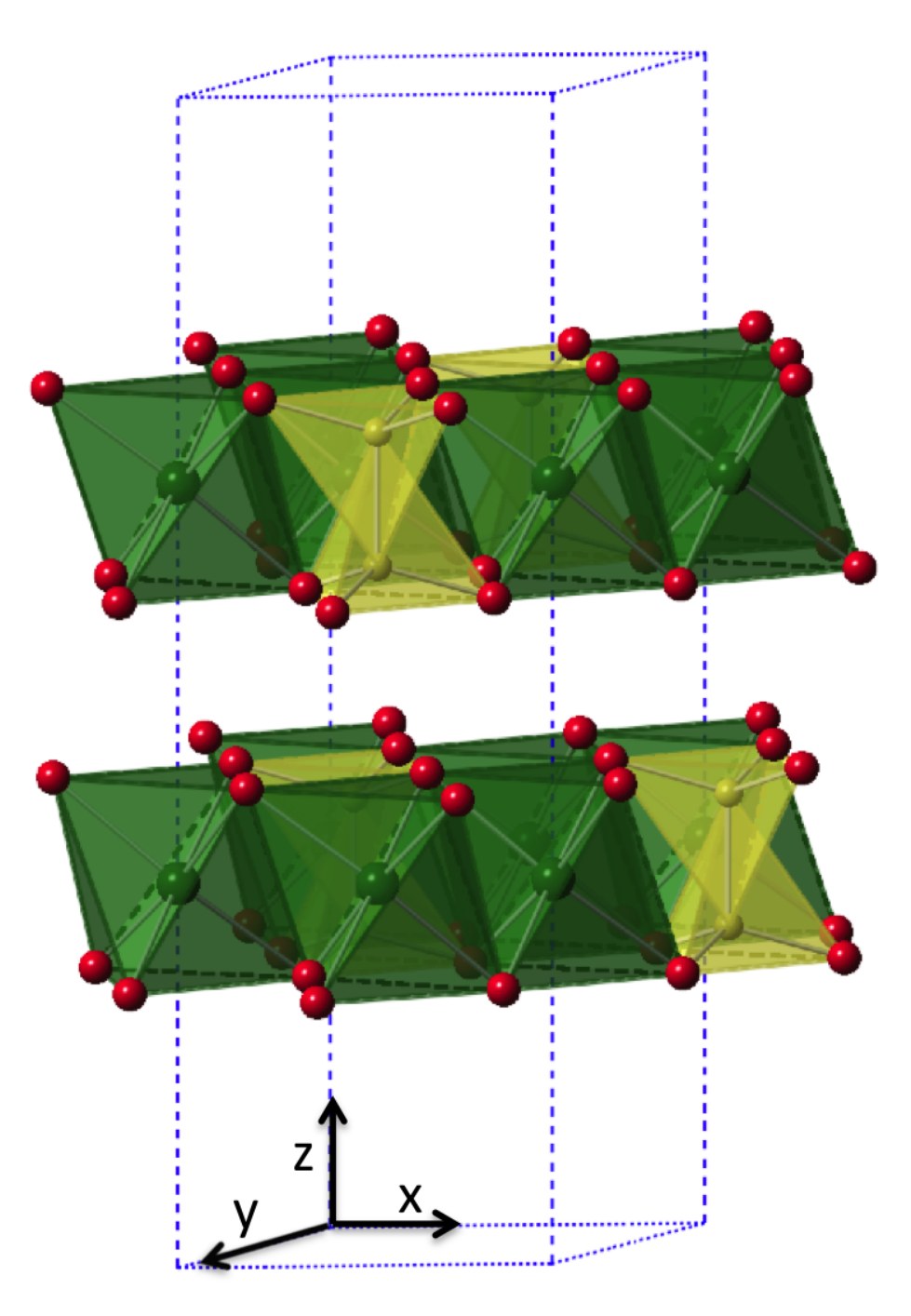}
\caption{The structure of $M_{2}$P$_{2}$Se$_{6}$ in hexagonal $R3$
symmetry has 2/3 of the octahedral sites filled with $M$ (green) and
the remaining 1/3 filled with a P$_{2}$ unit (gold).}
\label{fig:figure5}
\end{figure}

The rhombohedral $R\bar{3}$ and $R3$ structures
(Figure~\ref{fig:figure5}) reported for the $M_{2}$P$_{2}X_{6}$
entries are derived from the CdCl$_{2}$ structure by replacing the
halogen with S or Se, occupying 2/3 of the octahedral sites with $M$
and the remaining one-third by a diphosphorous unit aligned along the
$c$ direction~\cite{Boucher95p952}. The polar distortion in $R3$ is a
$\Gamma_{1}^{-}$ mode which consists of small displacements of the $M$
and P$_{2}$ units along the $c$ direction.

We used a first-principles approach with LSDA to investigate possible
ferroelectricity in rhombohedral both Cd$_{2}$P$_{2}$S$_{6}$ and
Fe$_{2}$P$_{2}$Se$_{6}$. With the reported $R3$ structures as the
starting structure, we performed structural relaxations to find that
the polar distortion disappears and the system relaxes to a nonpolar
$R{\bar3}$ structure. While it might be that a more accurate treatment
of electronic correlations, for example with DFT+$U$, is needed to
capture the ferroelectric instability, from the present result we
conclude that neither Fe$_{2}$P$_{2}$Se$_{6}$ or
Cd$_{2}$P$_{2}$S$_{6}$ is ferroelectric. As in the case of
Sr(Sb$_{1/2}$Mn$_{1/2}$)O$_{3}$ discussed previosuly, this negative
result is disappointing in the search for new ferrroelectrics, but
does illustrate the value of including first principles results in
evaluating structural data.

While this investigation did not definitively identify any new
ferroelectric compounds, the results suggest that it is a useful
strategy to consider the variety of structure types exhibited by a
simple compositional formula (in this case, 1:1:3 with P and either S
or Se). As in this case, this approach could bring to light smaller
structural families which have received less attention, but could have
additional, as yet unknown, representatives with polar instabilities
and desirable properties.

\section{\label{sec:level1} Prospects}

First principles predictions of structures and properties of real and
hypothetical compounds are a powerful tool in the design of new
materials with desirable properties~\cite{Fennie06p267602,
Lee10p954}. This power can be greatly amplified by integration of
these methods, which address individual compounds one at a time, with
database analysis, which permits a global view of the structure and
stability of compounds and the relationships between compounds and
classes of compounds~\cite{Trimarchi07p104113, Zhang12p1425}. With
recent increasing acceptance of this idea, it has become clear that
there are many ways in which this integration can be achieved. In this
paper, we have described how we have identified various strategies for
this integration through three exploratory investigations searching
for new ferroelectric semiconductors, in systems ranging from the
familiar family of perovskite oxides with a main group element (Sb)
sharing the $B$ site with a transition metal (Mn) to promote a lower
band gap, to the same combination of elements (Sb, Mn) in an
edge-sharing octahedral structure, to chalcogenides based on
P$_{2}X_{6}$ clusters.  In this final section, we summarize the
general strategies that have emerged from our work so far, and the
main avenues we have identified for further investigation.

As a starting point, our survey of all polar systems in the ICSD
allowed us to consider a very wide range of systems as candidate
ferroelectrics. One result from the polar compound survey was the
large number of orthorhombic polar compounds. This offers the exciting
opportunity to perform a systematic investigation of these structures
and, with the help of first principles results, to understand the
underlying mechanisms that favor arrangement into polar structures,
especially those which are related by small polar distortions to
high-symmetry reference structures and thus are candidate
ferroelectrics. The elucidation of one or more of these mechanisms
would point the way to the design of related and perhaps radically new
ferroelectric compounds.

One issue identified in our survey of the polar compounds is the need
for better information about structure types. It would be very useful
and quite practicable to introduce an automated classification of
structure types within a given space group given occupied Wyckoff
positions and structural parameters, making it easier to understand
the structural information and to identify the relationships between
compounds.

To screen the full collection of polar systems for systems with low
band gap, we formulated a criterion that the system should include
main group cations bound to either oxygen, sulfur or selenium. Our
first investigation, described in Section 3.2, focused on a particular
compound that satisfied both this chemical criterion and the
structural criterion described above, and had in a previous database
search~\cite{Abrahams96p790} been identified as a candidate
ferroelectric. Our initial purpose was to use first principles methods
to more fully characterize the system, and in particular to obtain the
band gap. However, it proved that the structure is in fact nonpolar,
consistent with other reports in the literature. A similar conclusion
was reached in the investigation of the reported polar $R3$
chalcogenide Fe$_2$P$_2$Se$_6$ in Section 3.4. The lesson is that
misassignment of nonpolar structures as polar does occur, and can be
confirmed with the assistance of first principles results.

A complementary strategy emerged from the relationship of
ferroelectrics to high-symmetry nonpolar reference structures, and the
possibility that there might be many polar structures that have been
misassigned as nonpolar in the database.  Symmetry analysis tools such
as ISOTROPY allow the easy identification of the relationships between
different structure types in different space groups, particularly when
the space groups have a group-subgroup relationship and the low
symmetry structure is related by one or a few normal modes to the
high-symmetry structure. While proper ferroelectrics, related to a
high-symmetry reference structure by a zone-center polar mode, are
often relatively easy to identify even without symmetry analysis
software, this approach can also readily identify candidate improper
ferroelectrics, in which the symmetry analysis is more subtle. We
searched in particular for relatively low-symmetry nonpolar structure
types as candidate high-symmetry reference structures. In addition to
shifting the focus from the more familiar perovskites, this has the
advantage that improper ferroelectricity is more likely to be found in
relatively low symmetry nonpolar systems, in which a single nonpolar
mode, which need not be at the zone center, can suffice to break the
symmetry that forbids spontaneous polarization. For our second
investigation, described in Section 3.3, we chose a
particular system with a low-symmetry nonpolar structure,
schafarzikite $M$Sb$_{2}$O$_{4}$, and checked for previously
unrecognized instabilities of modes that would break the symmetry to a
polar structure, both at the zone center and at other wave vectors as
determined by symmetry analysis. Despite the lack of instabilities for
this particular case, we believe that similar investigations of a
wider range of systems will yield positive results.  This suggests the
usefulness of a complete compilation of a list of space groups and
modes at high-symmetry wavevectors that break symmetry in such a way
as to produce a polar space group.

The identification of polar instabilities in nonpolar reference
structures could be facilitated by a complementary approach based on
the first principles computation of full phonon dispersions.  In order
for a system to exhibit either proper or improper ferroelectricity, it
is necessary that the relevant mode of the high-symmetry structure be
unstable. The first principles computation of the phonon dispersion of
a large number of nonpolar systems would allow screening of phonon
instabilities or low-frequency (marginally stable) modes, leading to
the identification of actual or incipient proper or improper
ferroelectrics. More generally, the value of a systematic collection
of first-principles phonon information has long been known for
spectroscopic (IR/Raman) identification~\cite{Caracas11p437}.  With
respect to the misassignment of polar structures as nonpolar, or the
identification of low-temperature transitions, systematic information
about the phonon dispersions of nonpolar structures would establish
how prevalent these might be, and assist in the development of
principles to identify nonpolar structure types susceptible to polar
instabilities.

Another strategy for the use of the database is to use a particular
system of interest as a starting point and to survey compounds with
the same structure, first to establish a pattern, and then to fill out
the pattern to search for more and possibly better
representatives. This was our initial purpose in Section 3.4, starting
from Sn$_{2}$P$_{2}$S$_{6}$ and Pb$_{2}$P$_{2}$S$_{6}$, though in this
case our search was unsuccessful. More generally, the capability of
first principles calculations to compute the structure and stability
of compounds lends itself naturally to high-throughput studies, in
which a large number of compounds with a common formula and structure
can be systematically searched for local stability, band gap, and
functional properties such as elastic, dielectric and piezoelectric
responses. Recent examples of such studies, discussed in the
Introduction, include our own study of ferroelectricity and
piezoelectricity in half-Heusler compounds~\cite{Roy11preprint}. As an
outgrowth of that project, we subsequently considered the six common
structure types with the same 1:1:1 formula, finding a class of
previously overlooked ferroelectrics in the LiGaGe structure type
(space group 186) and identifying individual examples in a
high-throughput study~\cite{Bennett12preprint_fe}. This demonstrates
the potential of this strategy in bringing to light structural
families which have received less attention, but could have
additional, as yet unknown, representatives with polar instabilities
and desirable properties.

With increasingly powerful first principles approaches and with
dramatic improvements in databases and in the synthetic capabilities
needed to realize new materials in the laboratory, this is truly an
exciting time for materials design. We eagerly look forward to rapid
progress in this field and the successful discovery and design of new
materials with fundamental scientific interest and valuable
technological applications.

\vspace{0.3cm}
\noindent{\bf Acknowledgments}
\vspace{0.3cm}
 
This work was supported by ONR grants N00014-09-1-0300 and MURI ARO
Grant W911NF-07-1-0410. We thank V. R. Cooper, P. K. Davies,
C. J. Fennie, S. P. Halasyamani, D. R. Hamann, S. E. Mason,
A. M. Rappe, A. Roy, J. F. Scott, R. Seshadri and D. Vanderbilt for
useful discussions.  K. M. R. also acknowledges the Aspen Center for
Physics and the hospitality of the Materials Department at University
of California, Santa Barbara, where part of this work was performed.

\bibliography{thebibliography}

\begin{thebibliography}{83}
\expandafter\ifx\csname natexlab\endcsname\relax\def\natexlab#1{#1}\fi
\expandafter\ifx\csname bibnamefont\endcsname\relax
  \def\bibnamefont#1{#1}\fi
\expandafter\ifx\csname bibfnamefont\endcsname\relax
  \def\bibfnamefont#1{#1}\fi
\expandafter\ifx\csname citenamefont\endcsname\relax
  \def\citenamefont#1{#1}\fi
\expandafter\ifx\csname url\endcsname\relax
  \def\url#1{\texttt{#1}}\fi
\expandafter\ifx\csname urlprefix\endcsname\relax\def\urlprefix{URL }\fi
\providecommand{\bibinfo}[2]{#2}
\providecommand{\eprint}[2][]{\url{#2}}

\bibitem[{\citenamefont{Fennie and Rabe}(2006)}]{Fennie06p267602}
\bibinfo{author}{\bibfnamefont{C.~J.} \bibnamefont{Fennie}} \bibnamefont{and}
  \bibinfo{author}{\bibfnamefont{K.~M.} \bibnamefont{Rabe}},
  \bibinfo{journal}{Phys. Rev. Lett.} \textbf{\bibinfo{volume}{97}},
  \bibinfo{pages}{267602} (\bibinfo{year}{2006}).

\bibitem[{\citenamefont{Lee et~al.}(2010)\citenamefont{Lee, Fang, Vlahos, Ke,
  Jung, Kourkoutis, Kim, Ryan, Heeg, Roeckerath et~al.}}]{Lee10p954}
\bibinfo{author}{\bibfnamefont{J.}~\bibnamefont{Lee}},
  \bibinfo{author}{\bibfnamefont{L.}~\bibnamefont{Fang}},
  \bibinfo{author}{\bibfnamefont{E.}~\bibnamefont{Vlahos}},
  \bibinfo{author}{\bibfnamefont{X.}~\bibnamefont{Ke}},
  \bibinfo{author}{\bibfnamefont{Y.}~\bibnamefont{Jung}},
  \bibinfo{author}{\bibfnamefont{L.~F.} \bibnamefont{Kourkoutis}},
  \bibinfo{author}{\bibfnamefont{J.-W.} \bibnamefont{Kim}},
  \bibinfo{author}{\bibfnamefont{P.}~\bibnamefont{Ryan}},
  \bibinfo{author}{\bibfnamefont{T.}~\bibnamefont{Heeg}},
  \bibinfo{author}{\bibfnamefont{M.}~\bibnamefont{Roeckerath}},
  \bibnamefont{et~al.}, \bibinfo{journal}{Nature Mater.}
  \textbf{\bibinfo{volume}{9}}, \bibinfo{pages}{954} (\bibinfo{year}{2010}).

\bibitem[{\citenamefont{Scott and
  Paz\hspace{5pt}de\hspace{5pt}Araujo}(1989)}]{Scott89p1400}
\bibinfo{author}{\bibfnamefont{J.~F.} \bibnamefont{Scott}} \bibnamefont{and}
  \bibinfo{author}{\bibfnamefont{C.~A.}
  \bibnamefont{Paz\hspace{5pt}de\hspace{5pt}Araujo}},
  \bibinfo{journal}{Science} \textbf{\bibinfo{volume}{246}},
  \bibinfo{pages}{1400} (\bibinfo{year}{1989}).

\bibitem[{\citenamefont{Izyumskaya et~al.}(2007)\citenamefont{Izyumskaya,
  Alivov, Cho, Morkoc, Lee, and Kang}}]{Izyumskaya07p111}
\bibinfo{author}{\bibfnamefont{N.}~\bibnamefont{Izyumskaya}},
  \bibinfo{author}{\bibfnamefont{Y.~I.} \bibnamefont{Alivov}},
  \bibinfo{author}{\bibfnamefont{S.~J.} \bibnamefont{Cho}},
  \bibinfo{author}{\bibfnamefont{H.}~\bibnamefont{Morkoc}},
  \bibinfo{author}{\bibfnamefont{H.}~\bibnamefont{Lee}}, \bibnamefont{and}
  \bibinfo{author}{\bibfnamefont{Y.~S.} \bibnamefont{Kang}},
  \bibinfo{journal}{Crit. Rev. in Solid State and Mater. Sci.}
  \textbf{\bibinfo{volume}{32}}, \bibinfo{pages}{111} (\bibinfo{year}{2007}).

\bibitem[{\citenamefont{Mannhart and Schlom}(2010)}]{Mannhart10p1608}
\bibinfo{author}{\bibfnamefont{J.}~\bibnamefont{Mannhart}} \bibnamefont{and}
  \bibinfo{author}{\bibfnamefont{D.~G.} \bibnamefont{Schlom}},
  \bibinfo{journal}{Science} \textbf{\bibinfo{volume}{327}},
  \bibinfo{pages}{1608} (\bibinfo{year}{2010}).

\bibitem[{\citenamefont{Clingman and Morore}(1961)}]{Clingman61p675}
\bibinfo{author}{\bibfnamefont{W.~H.} \bibnamefont{Clingman}} \bibnamefont{and}
  \bibinfo{author}{\bibfnamefont{R.~G.} \bibnamefont{Morore}},
  \bibinfo{journal}{J. Appl. Phys.} \textbf{\bibinfo{volume}{32}},
  \bibinfo{pages}{675} (\bibinfo{year}{1961}).

\bibitem[{\citenamefont{Edelstein}(1995)}]{Edelstein95p2004}
\bibinfo{author}{\bibfnamefont{V.~M.} \bibnamefont{Edelstein}},
  \bibinfo{journal}{Phys. Rev. Lett.} \textbf{\bibinfo{volume}{75}},
  \bibinfo{pages}{2004} (\bibinfo{year}{1995}).

\bibitem[{\citenamefont{Fridkin}(2001)}]{Fridkin01p654}
\bibinfo{author}{\bibfnamefont{V.~M.} \bibnamefont{Fridkin}},
  \bibinfo{journal}{Crystallog. Rep.} \textbf{\bibinfo{volume}{46}},
  \bibinfo{pages}{6754} (\bibinfo{year}{2001}).

\bibitem[{\citenamefont{Yang et~al.}(2009)\citenamefont{Yang, Martin, Byrnes,
  Conry, Basu, Paran, Reichertz, Ihlefeld, Adamo, Melville
  et~al.}}]{Yang09p062909}
\bibinfo{author}{\bibfnamefont{S.~Y.} \bibnamefont{Yang}},
  \bibinfo{author}{\bibfnamefont{L.~W.} \bibnamefont{Martin}},
  \bibinfo{author}{\bibfnamefont{S.~J.} \bibnamefont{Byrnes}},
  \bibinfo{author}{\bibfnamefont{T.~E.} \bibnamefont{Conry}},
  \bibinfo{author}{\bibfnamefont{S.~R.} \bibnamefont{Basu}},
  \bibinfo{author}{\bibfnamefont{D.}~\bibnamefont{Paran}},
  \bibinfo{author}{\bibfnamefont{L.}~\bibnamefont{Reichertz}},
  \bibinfo{author}{\bibfnamefont{J.}~\bibnamefont{Ihlefeld}},
  \bibinfo{author}{\bibfnamefont{C.}~\bibnamefont{Adamo}},
  \bibinfo{author}{\bibfnamefont{A.}~\bibnamefont{Melville}},
  \bibnamefont{et~al.}, \bibinfo{journal}{Appl. Phys. Lett.}
  \textbf{\bibinfo{volume}{95}}, \bibinfo{pages}{062909}
  (\bibinfo{year}{2009}).

\bibitem[{\citenamefont{Alexe and Hesse}(2011)}]{Alexe11p256}
\bibinfo{author}{\bibfnamefont{M.}~\bibnamefont{Alexe}} \bibnamefont{and}
  \bibinfo{author}{\bibfnamefont{D.}~\bibnamefont{Hesse}},
  \bibinfo{journal}{Nat. Commun.} \textbf{\bibinfo{volume}{2}},
  \bibinfo{pages}{256} (\bibinfo{year}{2011}).

\bibitem[{\citenamefont{Katiyar et~al.}(2011)\citenamefont{Katiyar, Kumar,
  Morell, Scott, and Katiyar}}]{Katiyar11p092906}
\bibinfo{author}{\bibfnamefont{R.~K.} \bibnamefont{Katiyar}},
  \bibinfo{author}{\bibfnamefont{A.}~\bibnamefont{Kumar}},
  \bibinfo{author}{\bibfnamefont{G.}~\bibnamefont{Morell}},
  \bibinfo{author}{\bibfnamefont{J.~F.} \bibnamefont{Scott}}, \bibnamefont{and}
  \bibinfo{author}{\bibfnamefont{R.~S.} \bibnamefont{Katiyar}},
  \bibinfo{journal}{Appl. Phys. Lett.} \textbf{\bibinfo{volume}{99}},
  \bibinfo{pages}{092906} (\bibinfo{year}{2011}).

\bibitem[{\citenamefont{Choi et~al.}(2009)\citenamefont{Choi, Lee, Choi,
  Kirukhin, and Cheong}}]{Choi09p63}
\bibinfo{author}{\bibfnamefont{T.}~\bibnamefont{Choi}},
  \bibinfo{author}{\bibfnamefont{S.}~\bibnamefont{Lee}},
  \bibinfo{author}{\bibfnamefont{Y.~J.} \bibnamefont{Choi}},
  \bibinfo{author}{\bibfnamefont{V.}~\bibnamefont{Kirukhin}}, \bibnamefont{and}
  \bibinfo{author}{\bibfnamefont{S.~W.} \bibnamefont{Cheong}},
  \bibinfo{journal}{Science} \textbf{\bibinfo{volume}{324}},
  \bibinfo{pages}{63} (\bibinfo{year}{2009}).

\bibitem[{\citenamefont{Wang et~al.}(2003)\citenamefont{Wang, Neaton, Zheng,
  Nagarajan, Ogale, Liu, Viehland, Vaithyanathan, Schlom, Waghmare
  et~al.}}]{Wang03p1719}
\bibinfo{author}{\bibfnamefont{J.}~\bibnamefont{Wang}},
  \bibinfo{author}{\bibfnamefont{J.~B.} \bibnamefont{Neaton}},
  \bibinfo{author}{\bibfnamefont{H.}~\bibnamefont{Zheng}},
  \bibinfo{author}{\bibfnamefont{V.}~\bibnamefont{Nagarajan}},
  \bibinfo{author}{\bibfnamefont{S.~B.} \bibnamefont{Ogale}},
  \bibinfo{author}{\bibfnamefont{B.}~\bibnamefont{Liu}},
  \bibinfo{author}{\bibfnamefont{D.}~\bibnamefont{Viehland}},
  \bibinfo{author}{\bibfnamefont{V.}~\bibnamefont{Vaithyanathan}},
  \bibinfo{author}{\bibfnamefont{D.~G.} \bibnamefont{Schlom}},
  \bibinfo{author}{\bibfnamefont{U.~V.} \bibnamefont{Waghmare}},
  \bibnamefont{et~al.}, \bibinfo{journal}{Science}
  \textbf{\bibinfo{volume}{299}}, \bibinfo{pages}{1719} (\bibinfo{year}{2003}),
  ISSN \bibinfo{issn}{0036-8075},
  \urlprefix\url{http://dx.doi.org/10.1126/science.1080615}.

\bibitem[{\citenamefont{Catalan and Scott}(2009)}]{Catalan09p2463}
\bibinfo{author}{\bibfnamefont{G.}~\bibnamefont{Catalan}} \bibnamefont{and}
  \bibinfo{author}{\bibfnamefont{J.~F.} \bibnamefont{Scott}},
  \bibinfo{journal}{Adv. Mater.} \textbf{\bibinfo{volume}{21}},
  \bibinfo{pages}{2463} (\bibinfo{year}{2009}).

\bibitem[{\citenamefont{Bennett et~al.}(2008)\citenamefont{Bennett, Grinberg,
  and Rappe}}]{Bennett08p17409}
\bibinfo{author}{\bibfnamefont{J.~W.} \bibnamefont{Bennett}},
  \bibinfo{author}{\bibfnamefont{I.}~\bibnamefont{Grinberg}}, \bibnamefont{and}
  \bibinfo{author}{\bibfnamefont{A.~M.} \bibnamefont{Rappe}},
  \bibinfo{journal}{J. Am. Chem. Soc.} \textbf{\bibinfo{volume}{130}},
  \bibinfo{pages}{17409} (\bibinfo{year}{2008}).

\bibitem[{\citenamefont{Gou et~al.}(2011)\citenamefont{Gou, Bennett, Takenaka,
  and Rappe}}]{Gou11p205115}
\bibinfo{author}{\bibfnamefont{G.~Y.} \bibnamefont{Gou}},
  \bibinfo{author}{\bibfnamefont{J.~W.} \bibnamefont{Bennett}},
  \bibinfo{author}{\bibfnamefont{H.}~\bibnamefont{Takenaka}}, \bibnamefont{and}
  \bibinfo{author}{\bibfnamefont{A.~M.} \bibnamefont{Rappe}},
  \bibinfo{journal}{Phys. Rev. B.} \textbf{\bibinfo{volume}{83}},
  \bibinfo{pages}{205115} (\bibinfo{year}{2011}).

\bibitem[{\citenamefont{Bennett et~al.}(2010)\citenamefont{Bennett, Grinberg,
  Davies, and Rappe}}]{Bennett10p184106}
\bibinfo{author}{\bibfnamefont{J.~W.} \bibnamefont{Bennett}},
  \bibinfo{author}{\bibfnamefont{I.}~\bibnamefont{Grinberg}},
  \bibinfo{author}{\bibfnamefont{P.~K.} \bibnamefont{Davies}},
  \bibnamefont{and} \bibinfo{author}{\bibfnamefont{A.~M.} \bibnamefont{Rappe}},
  \bibinfo{journal}{Phys. Rev. B.} \textbf{\bibinfo{volume}{82}},
  \bibinfo{pages}{184106} (\bibinfo{year}{2010}).

\bibitem[{\citenamefont{Qi et~al.}(2011)\citenamefont{Qi, Curnan, Kim, Bennett,
  Grinberg, and Rappe}}]{Qi11p245206}
\bibinfo{author}{\bibfnamefont{T.}~\bibnamefont{Qi}},
  \bibinfo{author}{\bibfnamefont{M.~T.} \bibnamefont{Curnan}},
  \bibinfo{author}{\bibfnamefont{S.}~\bibnamefont{Kim}},
  \bibinfo{author}{\bibfnamefont{J.~W.} \bibnamefont{Bennett}},
  \bibinfo{author}{\bibfnamefont{I.}~\bibnamefont{Grinberg}}, \bibnamefont{and}
  \bibinfo{author}{\bibfnamefont{A.~M.} \bibnamefont{Rappe}},
  \bibinfo{journal}{Phys. Rev. B.} p. \bibinfo{pages}{245206}
  (\bibinfo{year}{2011}).

\bibitem[{\citenamefont{Berger et~al.}(2011)\citenamefont{Berger, Fennie, and
  Neaton}}]{Berger11p146804}
\bibinfo{author}{\bibfnamefont{R.~F.} \bibnamefont{Berger}},
  \bibinfo{author}{\bibfnamefont{C.~J.} \bibnamefont{Fennie}},
  \bibnamefont{and} \bibinfo{author}{\bibfnamefont{J.~B.}
  \bibnamefont{Neaton}}, \bibinfo{journal}{Phys. Rev. Lett.}
  \textbf{\bibinfo{volume}{107}}, \bibinfo{pages}{146804}
  (\bibinfo{year}{2011}).

\bibitem[{\citenamefont{Haeni et~al.}(2004)\citenamefont{Haeni, Irvin, Chang,
  and et~al.}}]{Haeni04p758}
\bibinfo{author}{\bibfnamefont{J.~H.} \bibnamefont{Haeni}},
  \bibinfo{author}{\bibfnamefont{P.}~\bibnamefont{Irvin}},
  \bibinfo{author}{\bibfnamefont{W.}~\bibnamefont{Chang}}, \bibnamefont{and}
  \bibinfo{author}{\bibnamefont{et~al.}}, \bibinfo{journal}{Nature}
  \textbf{\bibinfo{volume}{430}}, \bibinfo{pages}{758} (\bibinfo{year}{2004}).

\bibitem[{\citenamefont{Eklund et~al.}(2009)\citenamefont{Eklund, Fennie, and
  Rabe}}]{Eklund09p220101R}
\bibinfo{author}{\bibfnamefont{C.~J.} \bibnamefont{Eklund}},
  \bibinfo{author}{\bibfnamefont{C.~J.} \bibnamefont{Fennie}},
  \bibnamefont{and} \bibinfo{author}{\bibfnamefont{K.~M.} \bibnamefont{Rabe}},
  \bibinfo{journal}{Phys. Rev. B.} \textbf{\bibinfo{volume}{79}},
  \bibinfo{pages}{220101} (\bibinfo{year}{2009}).

\bibitem[{\citenamefont{Picozzi et~al.}(2008)\citenamefont{Picozzi, Yamauchi,
  Sergienko, Sen, Senyal, and Dagotto}}]{Picozzi08p434208}
\bibinfo{author}{\bibfnamefont{S.}~\bibnamefont{Picozzi}},
  \bibinfo{author}{\bibfnamefont{K.}~\bibnamefont{Yamauchi}},
  \bibinfo{author}{\bibfnamefont{I.~A.} \bibnamefont{Sergienko}},
  \bibinfo{author}{\bibfnamefont{C.}~\bibnamefont{Sen}},
  \bibinfo{author}{\bibfnamefont{B.}~\bibnamefont{Senyal}}, \bibnamefont{and}
  \bibinfo{author}{\bibfnamefont{E.}~\bibnamefont{Dagotto}},
  \bibinfo{journal}{J. Phys.:Condens. Matter} \textbf{\bibinfo{volume}{20}},
  \bibinfo{pages}{434208} (\bibinfo{year}{2008}).

\bibitem[{\citenamefont{Fennie and Rabe}(2005)}]{Fennie05p100103R}
\bibinfo{author}{\bibfnamefont{C.~J.} \bibnamefont{Fennie}} \bibnamefont{and}
  \bibinfo{author}{\bibfnamefont{K.~M.} \bibnamefont{Rabe}},
  \bibinfo{journal}{Phys. Rev. B.} \textbf{\bibinfo{volume}{72}},
  \bibinfo{pages}{100103} (\bibinfo{year}{2005}).

\bibitem[{\citenamefont{Kim et~al.}(2010)\citenamefont{Kim, Koo, Sohn, and
  Shin}}]{Kim10p092902}
\bibinfo{author}{\bibfnamefont{J.}~\bibnamefont{Kim}},
  \bibinfo{author}{\bibfnamefont{Y.~M.} \bibnamefont{Koo}},
  \bibinfo{author}{\bibfnamefont{K.-S.} \bibnamefont{Sohn}}, \bibnamefont{and}
  \bibinfo{author}{\bibfnamefont{N.}~\bibnamefont{Shin}},
  \bibinfo{journal}{Appl. Phys. Lett.} \textbf{\bibinfo{volume}{97}},
  \bibinfo{pages}{092902} (\bibinfo{year}{2010}).

\bibitem[{\citenamefont{Benedek and Fennie}(2011)}]{Benedek11p107204}
\bibinfo{author}{\bibfnamefont{N.~A.} \bibnamefont{Benedek}} \bibnamefont{and}
  \bibinfo{author}{\bibfnamefont{C.~J.} \bibnamefont{Fennie}},
  \bibinfo{journal}{Phys. Rev. Lett.} \textbf{\bibinfo{volume}{106}},
  \bibinfo{pages}{107204} (\bibinfo{year}{2011}).

\bibitem[{\citenamefont{Fukushima et~al.}(2011)\citenamefont{Fukushima,
  stroppa, Picozzi, and Perez-Mato}}]{Fukushima11p12186}
\bibinfo{author}{\bibfnamefont{T.}~\bibnamefont{Fukushima}},
  \bibinfo{author}{\bibfnamefont{A.}~\bibnamefont{stroppa}},
  \bibinfo{author}{\bibfnamefont{S.}~\bibnamefont{Picozzi}}, \bibnamefont{and}
  \bibinfo{author}{\bibfnamefont{J.~M.} \bibnamefont{Perez-Mato}},
  \bibinfo{journal}{Phys. Chem. Chem. Phys.} \textbf{\bibinfo{volume}{13}},
  \bibinfo{pages}{12186} (\bibinfo{year}{2011}).

\bibitem[{\citenamefont{Bousquet et~al.}(2008)\citenamefont{Bousquet, Dawber,
  Stucki, Lichtensteiger, Hermet, Gariglio, Triscone, and
  Ghosez}}]{Bousquet08p732}
\bibinfo{author}{\bibfnamefont{E.}~\bibnamefont{Bousquet}},
  \bibinfo{author}{\bibfnamefont{M.}~\bibnamefont{Dawber}},
  \bibinfo{author}{\bibfnamefont{N.}~\bibnamefont{Stucki}},
  \bibinfo{author}{\bibfnamefont{C.}~\bibnamefont{Lichtensteiger}},
  \bibinfo{author}{\bibfnamefont{P.}~\bibnamefont{Hermet}},
  \bibinfo{author}{\bibfnamefont{S.}~\bibnamefont{Gariglio}},
  \bibinfo{author}{\bibfnamefont{J.-M.} \bibnamefont{Triscone}},
  \bibnamefont{and} \bibinfo{author}{\bibfnamefont{P.}~\bibnamefont{Ghosez}},
  \bibinfo{journal}{Nature} \textbf{\bibinfo{volume}{452}},
  \bibinfo{pages}{732} (\bibinfo{year}{2008}).

\bibitem[{\citenamefont{Maggard et~al.}(2003)\citenamefont{Maggard, Nault,
  Stern, and Poeppelmeier}}]{Maggard03p27}
\bibinfo{author}{\bibfnamefont{P.~A.} \bibnamefont{Maggard}},
  \bibinfo{author}{\bibfnamefont{T.~S.} \bibnamefont{Nault}},
  \bibinfo{author}{\bibfnamefont{C.~L.} \bibnamefont{Stern}}, \bibnamefont{and}
  \bibinfo{author}{\bibfnamefont{K.~R.} \bibnamefont{Poeppelmeier}},
  \bibinfo{journal}{J. Solis State Chem.} \textbf{\bibinfo{volume}{175}},
  \bibinfo{pages}{27} (\bibinfo{year}{2003}).

\bibitem[{\citenamefont{Marvel et~al.}(2007)\citenamefont{Marvel, Lesage, Baek,
  Halasyamani, Stern, and Poeppelmeier}}]{Marvel07p13963}
\bibinfo{author}{\bibfnamefont{M.~R.} \bibnamefont{Marvel}},
  \bibinfo{author}{\bibfnamefont{J.}~\bibnamefont{Lesage}},
  \bibinfo{author}{\bibfnamefont{J.}~\bibnamefont{Baek}},
  \bibinfo{author}{\bibfnamefont{P.~S.} \bibnamefont{Halasyamani}},
  \bibinfo{author}{\bibfnamefont{C.~L.} \bibnamefont{Stern}}, \bibnamefont{and}
  \bibinfo{author}{\bibfnamefont{K.~R.} \bibnamefont{Poeppelmeier}},
  \bibinfo{journal}{J. Amer. Chem. Soc.} \textbf{\bibinfo{volume}{129}},
  \bibinfo{pages}{13963} (\bibinfo{year}{2007}).

\bibitem[{\citenamefont{Chang et~al.}(2008)\citenamefont{Chang, Sivakumar, Ok,
  and Halasyamani}}]{Chang08p8511}
\bibinfo{author}{\bibfnamefont{H.~Y.} \bibnamefont{Chang}},
  \bibinfo{author}{\bibfnamefont{T.}~\bibnamefont{Sivakumar}},
  \bibinfo{author}{\bibfnamefont{K.~M.} \bibnamefont{Ok}}, \bibnamefont{and}
  \bibinfo{author}{\bibfnamefont{P.~S.} \bibnamefont{Halasyamani}},
  \bibinfo{journal}{Inorg. Chem.} \textbf{\bibinfo{volume}{47}},
  \bibinfo{pages}{8511} (\bibinfo{year}{2008}).

\bibitem[{\citenamefont{Kim et~al.}(2009)\citenamefont{Kim, Yeon, and
  Halasyamani}}]{Kim09p5335}
\bibinfo{author}{\bibfnamefont{S.-H.} \bibnamefont{Kim}},
  \bibinfo{author}{\bibfnamefont{J.}~\bibnamefont{Yeon}}, \bibnamefont{and}
  \bibinfo{author}{\bibfnamefont{P.~S.} \bibnamefont{Halasyamani}},
  \bibinfo{journal}{Chem. Mater.} \textbf{\bibinfo{volume}{21}},
  \bibinfo{pages}{5335} (\bibinfo{year}{2009}).

\bibitem[{\citenamefont{Chang et~al.}(2009)\citenamefont{Chang, Kim, Ok, and
  Halasyamani}}]{Chang09p6865}
\bibinfo{author}{\bibfnamefont{H.~Y.} \bibnamefont{Chang}},
  \bibinfo{author}{\bibfnamefont{S.-H.} \bibnamefont{Kim}},
  \bibinfo{author}{\bibfnamefont{K.~M.} \bibnamefont{Ok}}, \bibnamefont{and}
  \bibinfo{author}{\bibfnamefont{P.~S.} \bibnamefont{Halasyamani}},
  \bibinfo{journal}{J. Am. Chem. Soc.} \textbf{\bibinfo{volume}{131}},
  \bibinfo{pages}{6865} (\bibinfo{year}{2009}).

\bibitem[{\citenamefont{Belsky et~al.}(2002)\citenamefont{Belsky, Hellenbrandt,
  Karen, and Luksch}}]{Belsky02p364}
\bibinfo{author}{\bibfnamefont{A.}~\bibnamefont{Belsky}},
  \bibinfo{author}{\bibfnamefont{M.}~\bibnamefont{Hellenbrandt}},
  \bibinfo{author}{\bibfnamefont{V.~L.} \bibnamefont{Karen}}, \bibnamefont{and}
  \bibinfo{author}{\bibfnamefont{P.}~\bibnamefont{Luksch}},
  \bibinfo{journal}{Acta Cryst.} \textbf{\bibinfo{volume}{B58}},
  \bibinfo{pages}{364} (\bibinfo{year}{2002}).

\bibitem[{\citenamefont{Atuchin et~al.}(2004)\citenamefont{Atuchin, Kidyarov,
  and Pervukhina}}]{Atuchin04p411}
\bibinfo{author}{\bibfnamefont{V.~V.} \bibnamefont{Atuchin}},
  \bibinfo{author}{\bibfnamefont{B.~I.} \bibnamefont{Kidyarov}},
  \bibnamefont{and} \bibinfo{author}{\bibfnamefont{N.~V.}
  \bibnamefont{Pervukhina}}, \bibinfo{journal}{Comp. Mater. Sci.}
  \textbf{\bibinfo{volume}{30}}, \bibinfo{pages}{411} (\bibinfo{year}{2004}).

\bibitem[{\citenamefont{Halasyamani and
  Poeppelmeier}(1998)}]{Halasyamani98p2753}
\bibinfo{author}{\bibfnamefont{P.~S.} \bibnamefont{Halasyamani}}
  \bibnamefont{and} \bibinfo{author}{\bibfnamefont{K.~R.}
  \bibnamefont{Poeppelmeier}}, \bibinfo{journal}{Chem. Mater.}
  \textbf{\bibinfo{volume}{10}}, \bibinfo{pages}{2753} (\bibinfo{year}{1998}).

\bibitem[{\citenamefont{Abrahams}(1988)}]{Abrahams88p585}
\bibinfo{author}{\bibfnamefont{S.~C.} \bibnamefont{Abrahams}},
  \bibinfo{journal}{Acta. Cryst.} \textbf{\bibinfo{volume}{B44}},
  \bibinfo{pages}{585} (\bibinfo{year}{1988}).

\bibitem[{\citenamefont{Abrahams}(1996)}]{Abrahams96p790}
\bibinfo{author}{\bibfnamefont{S.~C.} \bibnamefont{Abrahams}},
  \bibinfo{journal}{Acta. Cryst.} \textbf{\bibinfo{volume}{B52}},
  \bibinfo{pages}{790} (\bibinfo{year}{1996}).

\bibitem[{\citenamefont{Abrahams}(2006)}]{Abrahams06p26}
\bibinfo{author}{\bibfnamefont{S.~C.} \bibnamefont{Abrahams}},
  \bibinfo{journal}{Acta. Cryst.} \textbf{\bibinfo{volume}{B62}},
  \bibinfo{pages}{26} (\bibinfo{year}{2006}).

\bibitem[{\citenamefont{Morgan et~al.}(2005)\citenamefont{Morgan, Ceder, and
  Curtarolo}}]{Morgan05p296}
\bibinfo{author}{\bibfnamefont{D.}~\bibnamefont{Morgan}},
  \bibinfo{author}{\bibfnamefont{G.}~\bibnamefont{Ceder}}, \bibnamefont{and}
  \bibinfo{author}{\bibfnamefont{S.}~\bibnamefont{Curtarolo}},
  \bibinfo{journal}{Mat. Sci. Technol.} \textbf{\bibinfo{volume}{16}},
  \bibinfo{pages}{296} (\bibinfo{year}{2005}).

\bibitem[{\citenamefont{Jain et~al.}(2011)\citenamefont{Jain, Hautier, Moore,
  Ong, Fischer, Mueller, Persson, and Ceder}}]{Jain11p2295}
\bibinfo{author}{\bibfnamefont{A.}~\bibnamefont{Jain}},
  \bibinfo{author}{\bibfnamefont{G.}~\bibnamefont{Hautier}},
  \bibinfo{author}{\bibfnamefont{C.~J.} \bibnamefont{Moore}},
  \bibinfo{author}{\bibfnamefont{S.~P.} \bibnamefont{Ong}},
  \bibinfo{author}{\bibfnamefont{C.~C.} \bibnamefont{Fischer}},
  \bibinfo{author}{\bibfnamefont{T.}~\bibnamefont{Mueller}},
  \bibinfo{author}{\bibfnamefont{K.~A.} \bibnamefont{Persson}},
  \bibnamefont{and} \bibinfo{author}{\bibfnamefont{G.}~\bibnamefont{Ceder}},
  \bibinfo{journal}{Comp. Mater. Science} \textbf{\bibinfo{volume}{50}},
  \bibinfo{pages}{2295} (\bibinfo{year}{2011}).

\bibitem[{\citenamefont{Fischer et~al.}(2006)\citenamefont{Fischer, Tibbetts,
  Morgan, and Ceder}}]{Fischer06p641}
\bibinfo{author}{\bibfnamefont{C.~C.} \bibnamefont{Fischer}},
  \bibinfo{author}{\bibfnamefont{K.~J.} \bibnamefont{Tibbetts}},
  \bibinfo{author}{\bibfnamefont{D.}~\bibnamefont{Morgan}}, \bibnamefont{and}
  \bibinfo{author}{\bibfnamefont{G.}~\bibnamefont{Ceder}},
  \bibinfo{journal}{Nature Mater.} \textbf{\bibinfo{volume}{5}},
  \bibinfo{pages}{641} (\bibinfo{year}{2006}).

\bibitem[{\citenamefont{Hautier et~al.}(2010)\citenamefont{Hautier, Fischer,
  Jain, Mueller, and Ceder}}]{Hautier10p3762}
\bibinfo{author}{\bibfnamefont{G.}~\bibnamefont{Hautier}},
  \bibinfo{author}{\bibfnamefont{C.~C.} \bibnamefont{Fischer}},
  \bibinfo{author}{\bibfnamefont{A.}~\bibnamefont{Jain}},
  \bibinfo{author}{\bibfnamefont{T.}~\bibnamefont{Mueller}}, \bibnamefont{and}
  \bibinfo{author}{\bibfnamefont{G.}~\bibnamefont{Ceder}},
  \bibinfo{journal}{Chem. Mater.} \textbf{\bibinfo{volume}{22}},
  \bibinfo{pages}{3762} (\bibinfo{year}{2010}).

\bibitem[{\citenamefont{Armiento et~al.}(2011)\citenamefont{Armiento, Kozinsky,
  Fornari, and Ceder}}]{Armiento11p014103}
\bibinfo{author}{\bibfnamefont{R.}~\bibnamefont{Armiento}},
  \bibinfo{author}{\bibfnamefont{B.}~\bibnamefont{Kozinsky}},
  \bibinfo{author}{\bibfnamefont{M.}~\bibnamefont{Fornari}}, \bibnamefont{and}
  \bibinfo{author}{\bibfnamefont{G.}~\bibnamefont{Ceder}},
  \bibinfo{journal}{Phys. Rev. B.} \textbf{\bibinfo{volume}{84}},
  \bibinfo{pages}{014103} (\bibinfo{year}{2011}).

\bibitem[{\citenamefont{Trimarchi and Zunger}(2007)}]{Trimarchi07p104113}
\bibinfo{author}{\bibfnamefont{G.}~\bibnamefont{Trimarchi}} \bibnamefont{and}
  \bibinfo{author}{\bibfnamefont{A.}~\bibnamefont{Zunger}},
  \bibinfo{journal}{Phys. Rev. B.} \textbf{\bibinfo{volume}{75}},
  \bibinfo{pages}{104113} (\bibinfo{year}{2007}).

\bibitem[{\citenamefont{Roy et~al.}(2012)\citenamefont{Roy, Bennett, Rabe, and
  Vanderbilt}}]{Roy11preprint}
\bibinfo{author}{\bibfnamefont{A.}~\bibnamefont{Roy}},
  \bibinfo{author}{\bibfnamefont{J.~W.} \bibnamefont{Bennett}},
  \bibinfo{author}{\bibfnamefont{K.~M.} \bibnamefont{Rabe}}, \bibnamefont{and}
  \bibinfo{author}{\bibfnamefont{D.}~\bibnamefont{Vanderbilt}},
  \bibinfo{journal}{Phys. Rev. Lett.} p. \bibinfo{pages}{under review}
  (\bibinfo{year}{2012}).

\bibitem[{\citenamefont{Bennett
  et~al.}(2012{\natexlab{a}})\citenamefont{Bennett, Garrity, Rabe, and
  Vanderbilt}}]{Bennett12preprint_fe}
\bibinfo{author}{\bibfnamefont{J.~W.} \bibnamefont{Bennett}},
  \bibinfo{author}{\bibfnamefont{K.~F.} \bibnamefont{Garrity}},
  \bibinfo{author}{\bibfnamefont{K.~M.} \bibnamefont{Rabe}}, \bibnamefont{and}
  \bibinfo{author}{\bibfnamefont{D.}~\bibnamefont{Vanderbilt}}, p.
  \bibinfo{pages}{in preparation} (\bibinfo{year}{2012}{\natexlab{a}}).

\bibitem[{\citenamefont{Bennett
  et~al.}(2012{\natexlab{b}})\citenamefont{Bennett, Zhang, Li, and
  Rabe}}]{Bennett12preprint_opto}
\bibinfo{author}{\bibfnamefont{J.~W.} \bibnamefont{Bennett}},
  \bibinfo{author}{\bibfnamefont{R.}~\bibnamefont{Zhang}},
  \bibinfo{author}{\bibfnamefont{J.}~\bibnamefont{Li}}, \bibnamefont{and}
  \bibinfo{author}{\bibfnamefont{K.~M.} \bibnamefont{Rabe}}, p.
  \bibinfo{pages}{in preparation} (\bibinfo{year}{2012}{\natexlab{b}}).

\bibitem[{\citenamefont{Yin and Cohen}(1982)}]{Yin82p5668}
\bibinfo{author}{\bibfnamefont{M.~T.} \bibnamefont{Yin}} \bibnamefont{and}
  \bibinfo{author}{\bibfnamefont{M.}~\bibnamefont{Cohen}},
  \bibinfo{journal}{Phys. Rev. B.} \textbf{\bibinfo{volume}{26}},
  \bibinfo{pages}{5668} (\bibinfo{year}{1982}).

\bibitem[{\citenamefont{Gonze et~al.}(2009)\citenamefont{Gonze, Amadon,
  Anglade, Beuken, Bottin, Boulanger, Bruneval, Caliste, Caracas, Cote
  et~al.}}]{Gonze09p2582}
\bibinfo{author}{\bibfnamefont{X.}~\bibnamefont{Gonze}},
  \bibinfo{author}{\bibfnamefont{B.}~\bibnamefont{Amadon}},
  \bibinfo{author}{\bibfnamefont{P.}~\bibnamefont{Anglade}},
  \bibinfo{author}{\bibfnamefont{J.~M.} \bibnamefont{Beuken}},
  \bibinfo{author}{\bibfnamefont{F.}~\bibnamefont{Bottin}},
  \bibinfo{author}{\bibfnamefont{P.}~\bibnamefont{Boulanger}},
  \bibinfo{author}{\bibfnamefont{F.}~\bibnamefont{Bruneval}},
  \bibinfo{author}{\bibfnamefont{D.}~\bibnamefont{Caliste}},
  \bibinfo{author}{\bibfnamefont{R.}~\bibnamefont{Caracas}},
  \bibinfo{author}{\bibfnamefont{M.}~\bibnamefont{Cote}}, \bibnamefont{et~al.},
  \bibinfo{journal}{Comp. Phys. Comm.} \textbf{\bibinfo{volume}{180}},
  \bibinfo{pages}{2582} (\bibinfo{year}{2009}).

\bibitem[{\citenamefont{Ramer and Rappe}(1999)}]{Ramer99p12471}
\bibinfo{author}{\bibfnamefont{N.~J.} \bibnamefont{Ramer}} \bibnamefont{and}
  \bibinfo{author}{\bibfnamefont{A.~M.} \bibnamefont{Rappe}},
  \bibinfo{journal}{Phys. Rev. B} \textbf{\bibinfo{volume}{59}},
  \bibinfo{pages}{12471} (\bibinfo{year}{1999}).

\bibitem[{\citenamefont{Rappe et~al.}(1990)\citenamefont{Rappe, Rabe, Kaxiras,
  and Joannopoulos}}]{Rappe90p1227}
\bibinfo{author}{\bibfnamefont{A.~M.} \bibnamefont{Rappe}},
  \bibinfo{author}{\bibfnamefont{K.~M.} \bibnamefont{Rabe}},
  \bibinfo{author}{\bibfnamefont{E.}~\bibnamefont{Kaxiras}}, \bibnamefont{and}
  \bibinfo{author}{\bibfnamefont{J.~D.} \bibnamefont{Joannopoulos}},
  \bibinfo{journal}{Phys. Rev. B Rapid Comm.} \textbf{\bibinfo{volume}{41}},
  \bibinfo{pages}{1227} (\bibinfo{year}{1990}).

\bibitem[{Opi()}]{Opium}
\bibinfo{howpublished}{http://opium.sourceforge.net}.

\bibitem[{\citenamefont{Monkhorst and Pack}(1976)}]{Monkhorst76p5188}
\bibinfo{author}{\bibfnamefont{H.~J.} \bibnamefont{Monkhorst}}
  \bibnamefont{and} \bibinfo{author}{\bibfnamefont{J.~D.} \bibnamefont{Pack}},
  \bibinfo{journal}{Phys. Rev. B} \textbf{\bibinfo{volume}{13}},
  \bibinfo{pages}{5188} (\bibinfo{year}{1976}).

\bibitem[{\citenamefont{Lima{}De{}Faria
  et~al.}(1990)\citenamefont{Lima{}De{}Faria, Hellner, Liebau, Makovicky, and
  Parthe}}]{LimaDeFaria90p1}
\bibinfo{author}{\bibfnamefont{J.}~\bibnamefont{Lima{}De{}Faria}},
  \bibinfo{author}{\bibfnamefont{E.}~\bibnamefont{Hellner}},
  \bibinfo{author}{\bibfnamefont{F.}~\bibnamefont{Liebau}},
  \bibinfo{author}{\bibfnamefont{E.}~\bibnamefont{Makovicky}},
  \bibnamefont{and} \bibinfo{author}{\bibfnamefont{E.}~\bibnamefont{Parthe}},
  \bibinfo{journal}{Acta. Cryst.} \textbf{\bibinfo{volume}{A46}},
  \bibinfo{pages}{1} (\bibinfo{year}{1990}).

\bibitem[{\citenamefont{Foster et~al.}(1996)\citenamefont{Foster, Nielson, and
  Abrahams}}]{Foster97p3076}
\bibinfo{author}{\bibfnamefont{M.~C.} \bibnamefont{Foster}},
  \bibinfo{author}{\bibfnamefont{R.~M.} \bibnamefont{Nielson}},
  \bibnamefont{and} \bibinfo{author}{\bibfnamefont{S.~C.}
  \bibnamefont{Abrahams}}, \bibinfo{journal}{J. Appl. Phys.}
  \textbf{\bibinfo{volume}{82}}, \bibinfo{pages}{3076} (\bibinfo{year}{1996}).

\bibitem[{\citenamefont{Politova et~al.}(1991)\citenamefont{Politova, Kaleva,
  Danilenko, Chuprakov, ivanov, and Venevtsev}}]{Politova91p2017}
\bibinfo{author}{\bibfnamefont{E.~D.} \bibnamefont{Politova}},
  \bibinfo{author}{\bibfnamefont{G.~M.} \bibnamefont{Kaleva}},
  \bibinfo{author}{\bibfnamefont{I.~N.} \bibnamefont{Danilenko}},
  \bibinfo{author}{\bibfnamefont{V.~F.} \bibnamefont{Chuprakov}},
  \bibinfo{author}{\bibfnamefont{S.~A.} \bibnamefont{ivanov}},
  \bibnamefont{and} \bibinfo{author}{\bibfnamefont{Y.~N.}
  \bibnamefont{Venevtsev}}, \bibinfo{journal}{Inorg. Mater.}
  \textbf{\bibinfo{volume}{26}}, \bibinfo{pages}{2017} (\bibinfo{year}{1991}).

\bibitem[{\citenamefont{Lufaso et~al.}(2004)\citenamefont{Lufaso, Woodward, and
  Goldberger}}]{Lufaso04p1651}
\bibinfo{author}{\bibfnamefont{M.~W.} \bibnamefont{Lufaso}},
  \bibinfo{author}{\bibfnamefont{P.~M.} \bibnamefont{Woodward}},
  \bibnamefont{and}
  \bibinfo{author}{\bibfnamefont{J.}~\bibnamefont{Goldberger}},
  \bibinfo{journal}{J. Solid State Chem.} \textbf{\bibinfo{volume}{177}},
  \bibinfo{pages}{1651} (\bibinfo{year}{2004}).

\bibitem[{\citenamefont{Cheah et~al.}(2006)\citenamefont{Cheah, Saines, and
  Kennedy}}]{Cheah06p1775}
\bibinfo{author}{\bibfnamefont{M.}~\bibnamefont{Cheah}},
  \bibinfo{author}{\bibfnamefont{P.~J.} \bibnamefont{Saines}},
  \bibnamefont{and} \bibinfo{author}{\bibfnamefont{B.~J.}
  \bibnamefont{Kennedy}}, \bibinfo{journal}{J. Solid State Chem.}
  \textbf{\bibinfo{volume}{179}}, \bibinfo{pages}{1775} (\bibinfo{year}{2006}).

\bibitem[{\citenamefont{Mandal et~al.}(2008)\citenamefont{Mandal, Polavets,
  Croft, and Greenblatt}}]{Mandal08p2325}
\bibinfo{author}{\bibfnamefont{T.~K.} \bibnamefont{Mandal}},
  \bibinfo{author}{\bibfnamefont{V.~V.} \bibnamefont{Polavets}},
  \bibinfo{author}{\bibfnamefont{M.}~\bibnamefont{Croft}}, \bibnamefont{and}
  \bibinfo{author}{\bibfnamefont{M.}~\bibnamefont{Greenblatt}},
  \bibinfo{journal}{J. Solid State Chem.} \textbf{\bibinfo{volume}{181}},
  \bibinfo{pages}{2325} (\bibinfo{year}{2008}).

\bibitem[{\citenamefont{Ivanov et~al.}(2009)\citenamefont{Ivanov, Norblad,
  Tellgren, and Hewat}}]{Ivanov09p822}
\bibinfo{author}{\bibfnamefont{S.~A.} \bibnamefont{Ivanov}},
  \bibinfo{author}{\bibfnamefont{P.}~\bibnamefont{Norblad}},
  \bibinfo{author}{\bibfnamefont{R.}~\bibnamefont{Tellgren}}, \bibnamefont{and}
  \bibinfo{author}{\bibfnamefont{A.}~\bibnamefont{Hewat}},
  \bibinfo{journal}{Mater. Res. Bull.} \textbf{\bibinfo{volume}{44}},
  \bibinfo{pages}{822} (\bibinfo{year}{2009}).

\bibitem[{\citenamefont{Campbell et~al.}(2006)\citenamefont{Campbell, Stokes,
  Tanner, and Hatch}}]{Campbell06p607}
\bibinfo{author}{\bibfnamefont{B.~J.} \bibnamefont{Campbell}},
  \bibinfo{author}{\bibfnamefont{H.~T.} \bibnamefont{Stokes}},
  \bibinfo{author}{\bibfnamefont{D.~E.} \bibnamefont{Tanner}},
  \bibnamefont{and} \bibinfo{author}{\bibfnamefont{D.~M.} \bibnamefont{Hatch}},
  \bibinfo{journal}{J. Appl. Cryst.} \textbf{\bibinfo{volume}{39}},
  \bibinfo{pages}{607} (\bibinfo{year}{2006}).

\bibitem[{\citenamefont{Lee and Rabe}(2010)}]{Lee10p207204}
\bibinfo{author}{\bibfnamefont{J.~H.} \bibnamefont{Lee}} \bibnamefont{and}
  \bibinfo{author}{\bibfnamefont{K.~M.} \bibnamefont{Rabe}},
  \bibinfo{journal}{Phys. Rev. Lett.} \textbf{\bibinfo{volume}{104}},
  \bibinfo{pages}{207204} (\bibinfo{year}{2010}).

\bibitem[{JFS()}]{JFScott}
\bibinfo{howpublished}{private communication}.

\bibitem[{\citenamefont{Schmid and Ascher}(1974)}]{Schmid74p2697}
\bibinfo{author}{\bibfnamefont{H.}~\bibnamefont{Schmid}} \bibnamefont{and}
  \bibinfo{author}{\bibfnamefont{E.}~\bibnamefont{Ascher}},
  \bibinfo{journal}{J. Phys. C: Solid State Phys.}
  \textbf{\bibinfo{volume}{7}}, \bibinfo{pages}{2697} (\bibinfo{year}{1974}).

\bibitem[{\citenamefont{Fennie et~al.}(2007)\citenamefont{Fennie, Seshadri, and
  Rabe}}]{Fennie07preprint}
\bibinfo{author}{\bibfnamefont{C.~J.} \bibnamefont{Fennie}},
  \bibinfo{author}{\bibfnamefont{R.}~\bibnamefont{Seshadri}}, \bibnamefont{and}
  \bibinfo{author}{\bibfnamefont{K.~M.} \bibnamefont{Rabe}},
  \bibinfo{journal}{arXiv:0712.1846}  (\bibinfo{year}{2007}).

\bibitem[{\citenamefont{Rabe}(2010)}]{Rabe10p211}
\bibinfo{author}{\bibfnamefont{K.~M.} \bibnamefont{Rabe}},
  \bibinfo{journal}{Ann. Rev. of Condend. Matter Phys.}
  \textbf{\bibinfo{volume}{1}}, \bibinfo{pages}{211} (\bibinfo{year}{2010}).

\bibitem[{\citenamefont{Whitaker et~al.}(2011)\citenamefont{Whitaker, Bayliss,
  Berry, and Greaves}}]{Whitaker11p14523}
\bibinfo{author}{\bibfnamefont{M.~J.} \bibnamefont{Whitaker}},
  \bibinfo{author}{\bibfnamefont{R.~D.} \bibnamefont{Bayliss}},
  \bibinfo{author}{\bibfnamefont{F.~J.} \bibnamefont{Berry}}, \bibnamefont{and}
  \bibinfo{author}{\bibfnamefont{C.}~\bibnamefont{Greaves}},
  \bibinfo{journal}{J. Mater. Chem.} \textbf{\bibinfo{volume}{21}},
  \bibinfo{pages}{14523} (\bibinfo{year}{2011}).

\bibitem[{\citenamefont{Puebla et~al.}(1982)\citenamefont{Puebla, Rios, Monge,
  and Rasines}}]{Puebla82p2020}
\bibinfo{author}{\bibfnamefont{E.~G.} \bibnamefont{Puebla}},
  \bibinfo{author}{\bibfnamefont{E.~G.} \bibnamefont{Rios}},
  \bibinfo{author}{\bibfnamefont{A.}~\bibnamefont{Monge}}, \bibnamefont{and}
  \bibinfo{author}{\bibfnamefont{I.}~\bibnamefont{Rasines}},
  \bibinfo{journal}{Acta. Crystallogr. B.} \textbf{\bibinfo{volume}{38}},
  \bibinfo{pages}{2020} (\bibinfo{year}{1982}).

\bibitem[{\citenamefont{Stokes et~al.}()\citenamefont{Stokes, Hatch, and
  Campbell}}]{ISOTROPY}
\bibinfo{author}{\bibfnamefont{H.~T.} \bibnamefont{Stokes}},
  \bibinfo{author}{\bibfnamefont{D.~M.} \bibnamefont{Hatch}}, \bibnamefont{and}
  \bibinfo{author}{\bibfnamefont{B.~J.} \bibnamefont{Campbell}},
  \bibinfo{howpublished}{http://stokes.byu.edu/isotropy.html}.

\bibitem[{\citenamefont{Bennett et~al.}(2009)\citenamefont{Bennett, Grinberg,
  and Rappe}}]{Bennett09p235115}
\bibinfo{author}{\bibfnamefont{J.~W.} \bibnamefont{Bennett}},
  \bibinfo{author}{\bibfnamefont{I.}~\bibnamefont{Grinberg}}, \bibnamefont{and}
  \bibinfo{author}{\bibfnamefont{A.~M.} \bibnamefont{Rappe}},
  \bibinfo{journal}{Phys. Rev. B.} \textbf{\bibinfo{volume}{79}},
  \bibinfo{pages}{235115} (\bibinfo{year}{2009}).

\bibitem[{\citenamefont{Moriya et~al.}(1998)\citenamefont{Moriya, Kuniyoshi,
  Tashita, Ozaki, Yano, and Matsuo}}]{Moriya98p3505}
\bibinfo{author}{\bibfnamefont{K.}~\bibnamefont{Moriya}},
  \bibinfo{author}{\bibfnamefont{H.}~\bibnamefont{Kuniyoshi}},
  \bibinfo{author}{\bibfnamefont{K.}~\bibnamefont{Tashita}},
  \bibinfo{author}{\bibfnamefont{Y.}~\bibnamefont{Ozaki}},
  \bibinfo{author}{\bibfnamefont{S.}~\bibnamefont{Yano}}, \bibnamefont{and}
  \bibinfo{author}{\bibfnamefont{T.}~\bibnamefont{Matsuo}},
  \bibinfo{journal}{J. Phys. Soc. Jap.} \textbf{\bibinfo{volume}{67}},
  \bibinfo{pages}{3505} (\bibinfo{year}{1998}).

\bibitem[{\citenamefont{van{}Loosdrecht
  et~al.}(1993)\citenamefont{van{}Loosdrecht, Maior, Molnar, Vysochanskii,
  van{}Bentum, and van{}Kempen}}]{vanLoosdrecht93p6014}
\bibinfo{author}{\bibfnamefont{P.~H.~M.} \bibnamefont{van{}Loosdrecht}},
  \bibinfo{author}{\bibfnamefont{M.~M.} \bibnamefont{Maior}},
  \bibinfo{author}{\bibfnamefont{S.~B.} \bibnamefont{Molnar}},
  \bibinfo{author}{\bibfnamefont{Y.~M.} \bibnamefont{Vysochanskii}},
  \bibinfo{author}{\bibfnamefont{P.~J.~M.} \bibnamefont{van{}Bentum}},
  \bibnamefont{and}
  \bibinfo{author}{\bibfnamefont{H.}~\bibnamefont{van{}Kempen}},
  \bibinfo{journal}{Phys. Rev. B.} \textbf{\bibinfo{volume}{48}},
  \bibinfo{pages}{6014} (\bibinfo{year}{1993}).

\bibitem[{\citenamefont{Eijt et~al.}(1998)\citenamefont{Eijt, Currat, Lorenzo,
  Saint-Gregoire, Katano, Janssen, Hennion, and Vysochanskii}}]{Eijt98p4811}
\bibinfo{author}{\bibfnamefont{S.~W.~H.} \bibnamefont{Eijt}},
  \bibinfo{author}{\bibfnamefont{R.}~\bibnamefont{Currat}},
  \bibinfo{author}{\bibfnamefont{J.~E.} \bibnamefont{Lorenzo}},
  \bibinfo{author}{\bibfnamefont{P.}~\bibnamefont{Saint-Gregoire}},
  \bibinfo{author}{\bibfnamefont{S.}~\bibnamefont{Katano}},
  \bibinfo{author}{\bibfnamefont{T.}~\bibnamefont{Janssen}},
  \bibinfo{author}{\bibfnamefont{B.}~\bibnamefont{Hennion}}, \bibnamefont{and}
  \bibinfo{author}{\bibfnamefont{Y.~M.} \bibnamefont{Vysochanskii}},
  \bibinfo{journal}{J. Phys.:Condens. Matter} \textbf{\bibinfo{volume}{10}},
  \bibinfo{pages}{4811} (\bibinfo{year}{1998}).

\bibitem[{\citenamefont{Maior et~al.}(1994)\citenamefont{Maior, Rasing, Eijt,
  van{}Loosdrecht, van{}Kempen, Molnar, Vysochanskii, Motrij, and
  Slivka}}]{Maior94p11211}
\bibinfo{author}{\bibfnamefont{M.~M.} \bibnamefont{Maior}},
  \bibinfo{author}{\bibfnamefont{T.}~\bibnamefont{Rasing}},
  \bibinfo{author}{\bibfnamefont{S.~W.~H.} \bibnamefont{Eijt}},
  \bibinfo{author}{\bibfnamefont{P.~H.~M.} \bibnamefont{van{}Loosdrecht}},
  \bibinfo{author}{\bibfnamefont{H.}~\bibnamefont{van{}Kempen}},
  \bibinfo{author}{\bibfnamefont{S.~B.} \bibnamefont{Molnar}},
  \bibinfo{author}{\bibfnamefont{Y.~M.} \bibnamefont{Vysochanskii}},
  \bibinfo{author}{\bibfnamefont{S.~F.} \bibnamefont{Motrij}},
  \bibnamefont{and} \bibinfo{author}{\bibfnamefont{V.~Y.}
  \bibnamefont{Slivka}}, \bibinfo{journal}{J. Phys.: Condens. Matter}
  \textbf{\bibinfo{volume}{6}}, \bibinfo{pages}{11211} (\bibinfo{year}{1994}).

\bibitem[{\citenamefont{Smirnov et~al.}(2000)\citenamefont{Smirnov, Hlinka, and
  Solov'ev}}]{Smirnov00p15051}
\bibinfo{author}{\bibfnamefont{M.~B.} \bibnamefont{Smirnov}},
  \bibinfo{author}{\bibfnamefont{J.}~\bibnamefont{Hlinka}}, \bibnamefont{and}
  \bibinfo{author}{\bibfnamefont{A.~V.} \bibnamefont{Solov'ev}},
  \bibinfo{journal}{Phys. Rev. B.} \textbf{\bibinfo{volume}{61}},
  \bibinfo{pages}{15051} (\bibinfo{year}{2000}).

\bibitem[{\citenamefont{Rushchanskii et~al.}(2007)\citenamefont{Rushchanskii,
  Vysochanskii, and Strauch}}]{Rushchanskii07p207601}
\bibinfo{author}{\bibfnamefont{K.~Z.} \bibnamefont{Rushchanskii}},
  \bibinfo{author}{\bibfnamefont{Y.~M.} \bibnamefont{Vysochanskii}},
  \bibnamefont{and} \bibinfo{author}{\bibfnamefont{D.}~\bibnamefont{Strauch}},
  \bibinfo{journal}{Phys. Rev. Lett.} \textbf{\bibinfo{volume}{99}},
  \bibinfo{pages}{207601} (\bibinfo{year}{2007}).

\bibitem[{\citenamefont{Cho et~al.}(2001)\citenamefont{Cho, Choi, and
  Vysochanskii}}]{Cho01p3317}
\bibinfo{author}{\bibfnamefont{Y.~W.} \bibnamefont{Cho}},
  \bibinfo{author}{\bibfnamefont{S.~K.} \bibnamefont{Choi}}, \bibnamefont{and}
  \bibinfo{author}{\bibfnamefont{Y.~M.} \bibnamefont{Vysochanskii}},
  \bibinfo{journal}{J. Mater. Res.} \textbf{\bibinfo{volume}{16}},
  \bibinfo{pages}{3317} (\bibinfo{year}{2001}).

\bibitem[{\citenamefont{Caracas and Gonze}(2002)}]{Caracas02p104106}
\bibinfo{author}{\bibfnamefont{R.}~\bibnamefont{Caracas}} \bibnamefont{and}
  \bibinfo{author}{\bibfnamefont{X.}~\bibnamefont{Gonze}},
  \bibinfo{journal}{Phys. Rev. B.} \textbf{\bibinfo{volume}{66}},
  \bibinfo{pages}{104106} (\bibinfo{year}{2002}).

\bibitem[{\citenamefont{Glukhov et~al.}(2011)\citenamefont{Glukhov, Fedyo, and
  Vysochanskii}}]{Glukhovpreprint}
\bibinfo{author}{\bibfnamefont{K.}~\bibnamefont{Glukhov}},
  \bibinfo{author}{\bibfnamefont{K.}~\bibnamefont{Fedyo}}, \bibnamefont{and}
  \bibinfo{author}{\bibfnamefont{Y.}~\bibnamefont{Vysochanskii}}, p.
  \bibinfo{pages}{arXiv:1108.2390v2} (\bibinfo{year}{2011}).

\bibitem[{\citenamefont{Boucher et~al.}(1995)\citenamefont{Boucher, Evain, and
  Brec}}]{Boucher95p952}
\bibinfo{author}{\bibfnamefont{F.}~\bibnamefont{Boucher}},
  \bibinfo{author}{\bibfnamefont{M.}~\bibnamefont{Evain}}, \bibnamefont{and}
  \bibinfo{author}{\bibfnamefont{R.}~\bibnamefont{Brec}},
  \bibinfo{journal}{Acta. Cryst.} \textbf{\bibinfo{volume}{B51}},
  \bibinfo{pages}{952} (\bibinfo{year}{1995}).

\bibitem[{\citenamefont{Ouvrard et~al.}(1985)\citenamefont{Ouvrard, Brec, and
  Rouxel}}]{Ouvrard85p1181}
\bibinfo{author}{\bibfnamefont{G.}~\bibnamefont{Ouvrard}},
  \bibinfo{author}{\bibfnamefont{R.}~\bibnamefont{Brec}}, \bibnamefont{and}
  \bibinfo{author}{\bibfnamefont{J.}~\bibnamefont{Rouxel}},
  \bibinfo{journal}{Mat. Res. Bull.} \textbf{\bibinfo{volume}{20}},
  \bibinfo{pages}{1181} (\bibinfo{year}{1985}).

\bibitem[{\citenamefont{Zhang et~al.}(2012)\citenamefont{Zhang, Yu, Zakutayev,
  and Zunger}}]{Zhang12p1425}
\bibinfo{author}{\bibfnamefont{X.}~\bibnamefont{Zhang}},
  \bibinfo{author}{\bibfnamefont{L.}~\bibnamefont{Yu}},
  \bibinfo{author}{\bibfnamefont{A.}~\bibnamefont{Zakutayev}},
  \bibnamefont{and} \bibinfo{author}{\bibfnamefont{A.}~\bibnamefont{Zunger}},
  \bibinfo{journal}{Adv. Funct. Mater.} \textbf{\bibinfo{volume}{22}},
  \bibinfo{pages}{1425} (\bibinfo{year}{2012}).

\bibitem[{\citenamefont{Caracas and Bobocioiu}(2011)}]{Caracas11p437}
\bibinfo{author}{\bibfnamefont{R.}~\bibnamefont{Caracas}} \bibnamefont{and}
  \bibinfo{author}{\bibfnamefont{E.}~\bibnamefont{Bobocioiu}},
  \bibinfo{journal}{American Mineralogist} \textbf{\bibinfo{volume}{96}},
  \bibinfo{pages}{437} (\bibinfo{year}{2011}).

\end{thebibliography}

\end{document}